# Retour sur les lingots de plomb de Comacchio (Ferrara, Italie) en passant par l'archéométrie et l'épigraphie

par

Claude DOMERGUE
UTAH, UMR 5608 CNRS
Maison de la Recherche, Université de Toulouse-le Mirail, 5 allée Antonio-Machado, 31058 Toulouse Cédex 09 France
Tél. 33 (0)5 61 86 89 74
claude.domergue@wanadoo.fr
Professeur émérite des Universités

Piero QUARATI
Dipartamento di Fisica, Politecnico di Torino et INFN, Sezione di Cagliari
Dipartimento di Fisica del Politecnico di Torino, corso Duca degli Abruzzi, 24 10129 Torino Italia
Tél. 39 011 5647315
piero.quarati@polito.it
Professeur de Physique Nucléaire

Antonio NESTA
Dipartamento di Fisica, Politecnico di Torino
(adrsse privée) Via Veracini 59 Firenze Italia
Tél. +39 348 0365301 ou 055 4233397
a_nesta@libero.it
Ingénieur Nucléaire

Pier Renato TRINCHERINI
Joint Research Centre of European Commission, Environment Institute, TP 290, Ispra (Va) Italy
galois23@libero.it`

Résumés

Jusqu'ici, le provenance des 102 lingots de plomb découverts en 1981 dans une épave à Comacchio (Ferrara, Italie), dans le delta du Pô, était incertaine et faisait l'objet d'hypothèses diverses. Récemment, la signature isotopique du plomb de vingt de ces lingots, représentatifs de l'ensemble du lot, a été déterminée au laboratoire du JRC de la Commission Européenne, à Ispra (Italie). L'examen des résultats de ces analyses suggère avec une très forte probabilité les mines de Carthagène-Mazarrón ou celles de la Sierra Almagrera, dans le sud-est de l'Espagne, comme lieu d'origine de ces lingots. La lecture ici proposée des timbres imprimés sur les lingots semble renforcer la première de ces deux provenances possibles, puisqu'on croit pouvoir identifier, entre autres noms, ceux de familles minières de *Carthago Noua* connues dans la première moitié du I$^{er}$ siècle avant J.-C.. Un autre timbre renferme le nom d'Agrippa, vraisemblablement le gendre d'Auguste. Le *terminus ante quem* des lingots serait alors 12 avant J.-C. et leur date se situerait entre –19 et –12. Le rôle des différents personnages mentionnés par les timbres est examiné dans une optique commerciale. Enfin on tente d'expliquer la présence de ce lot de lingots dans l'Adriatique-nord et de reconstituer l'itinéraire qu'ils ont suivi.

Hasta hoy, la procedencia de los 102 lingotes de plomo descubiertos en 1981 en un pecio cerca de Comacchio (Ferrara, Italia), en el delta del Po, estaba incierta ; se han emitido sobre esta cuestión varios hipótesis. Hace poco, se ha analizado la composición isotópica del plomo de veinte de estos lingotes, representativos del conjunto, en el laboratorio del JRC de la Comisión Europea de Ispra (Italia). El examen de los resultados indica con una gran probabilidad como región de origen de los lingotes, las minas de Cartagena-Mazarrón o las de la Sierra Almagrera, en el sureste de España. Ahora bien, se propone una lectura de los sellos impresos sobre los lingotes, la cual parece confirmar la primera de estas dos regiones, ya que se puede leer en algunos de ellos los nombres de unas familias mineras de *Carthago Noua,* ya conocidas en la primera mitad del siglo I a.C.. Otro sello lleva el nombre de Agrippa, que debe ser el yerno de Augusto. El *terminus ante quem* de los lingotes sería pues 12 a.C., y su fecha entre 19 y 12 a.C. El papel de las personas cuyos nombres aparecen en los sellos se estudia dentro de un marco comercial. Al final, se intenta explicar la presencia de este conjunto de lingotes en la parte norte del mar Adriático y reconstruir el itinerario seguido.

L'intérêt exceptionnel de l'épave romaine découverte légèrement au sud de l'actuel delta du Pô, à Comacchio, en 1981, a été bien mis en évidence par le magnifique catalogue de l'exposition, *Fortuna Maris*, qui, en 1990, a présenté au public à la fois l'épave et les marchandises qu'on y a découvertes, dans le cadre qui les accueillait définitivement, le palais Bellini à Comacchio[1]. La publication complète de cet ensemble n'est pas encore achevée, en revanche une partie du matériel métallique, les lingots de plomb (Fig. 1), a fait déjà l'objet de plusieurs études. Ainsi, F. Berti, directrice du Musée de Ferrara, en a donné la description en plusieurs occasions [2] : typologie, revue des dix timbres qu'ils portent, identification du personnage mentionné par l'un d'eux avec M. Vipsanius Agrippa. Par ailleurs, l'un de nous avait lui aussi identifié le principal personnage comme étant Agrippa, le gendre d'Auguste (63-12 av. J.-C), et cherché à retrouver, dans les autres timbres, les noms de personnes ayant pu faire partie de l'entourage de ce dernier, mais il avait buté sur la signification de quatre de ces timbres dont deux lui paraissaient représenter des symboles[3]. Enfin, M.P. García Bellido a donné de l'ensemble de ces timbres une interprétation militaire, qui, dans la production des lingots, fait intervenir, sous l'égide de M. Vipsanius Agrippa, des détachements des légions I, IIII Macedonica, X Gemina, qui avaient toutes été stationnées en Espagne lors de la conquête du nord-ouest, officiellement achevée par Agrippa lui-même en 19 avant J.-C.[4]. La question de la provenance de ces lingots a bien sûr été abordée par ces chercheurs : pour F. Berti, ils venaient d'Espagne[5], pour C. Domergue, plutôt des Balkans[6] ; pour M.P. García Bellido enfin, de La Serena, une région minière d'Estremadure[7]. Mais, quelle que fût la provenance envisagée, ils souhaitaient tous que des analyses fussent effectuées pour que pût être déterminée avec plus d'assurance la région d'origine[8].

Une première tentative eut lieu au début des années 1990 ; les premiers résultats indiquèrent trois origines possibles : Bulgarie, France, Espagne orientale[9], mais l'entreprise ne fut pas menée à son terme. Nous reprenons aujourd'hui la question sur la base d'analyses isotopiques du plomb de vingt lingots de Comacchio, parmi lesquels sont représentées les cinq séries distinguées par F. Berti. Les résultats de ces analyses seront présentés et discutés. On verra qu'ils suggèrent, comme région d'origine des lingots, le sud-est de l'Espagne, avec une grande probabilité en faveur des mines de *Carthago Noua* –Mazarrón et de celles de la Sierra Almagrera. Sur cette base, seront reprises la lecture et l'interprétation des timbres, de façon à situer ces lingots dans l'histoire de la production et du commerce du plomb hispanique[10].

+
+ +

1 - Les analyses

---

[1] Berti 1990.
[2] Berti 1985 , 1986, 1987.
[3] Domergue 1987 a (illustré de nombreuses photographies).
[4] Garcia Bellido 1994-1995, 1998 a et b ; García Bellido & Petac 1998.
[5] Berti 1985, 1986, 1987. Même provenance indiquée dans F. Berti, The Comacchio Wreck, dans *A Database on Ancient Ships (*http://www2.rgzm.de/navis/home/frames.htm).
[6] Domergue 1987 a, p. 139.
[7] García Bellido 1998 a, p. 36-37. Sur les mines antiques de plomb-argent de La Serena, voir Domergue 1987 b, p. 27-34 et fig. 73 ; 1990, carte 4.
[8] Berti 1987, p. 132 ; Domergue 1987 a, p. 141 ; García Bellido 1998 a, p. 36.
[9] Berti 1990, p. 76, n. 32.
[10] Nous ne revenons pas ici sur la description des lingots. Nous renvoyons pour cela à Domergue 1987 a et à Berti 1990. Nous conservons la division en cinq séries de F. Berti (1990, p. 74).

Vingt des cent-deux lingots de Comacchio ont été échantillonnés[11]. Les mesures des rapports isotopiques du plomb de ces vingt échantillons ont toutes été effectuées au laboratoire de Spectrométrie de Masse de l'European Joint Research Centre de Ispra (Italie), sur un spectromètre Finningan MAT 262-Thermal Ionization Mass Spectroscopy (TIMS), muni de multi-collecteurs de façon à recueillir et à mesurer jusqu'à cinq isotopes en même temps. Par ailleurs, quelque 300 autres échantillons soit de minerais provenant de mines antiques d'Espagne soit de lingots de plomb romains découverts dans l'aire méditerranéenne, ont été semblablement analysés par ce même laboratoire. À ces analyses s'ajoutent celles de minerais et de lingots du monde méditerranéen qui ont été publiées dans la littérature, si bien que nous disposons d'une banque de données bien nourrie, ce qui rend fiable la comparaison entre les données fournies par les vingt analyses des lingots de Comacchio et celles qui concernent les mines[12].

La précision des mesures est de 0,01% pour le rapport isotopique $^{206}Pb/^{207}Pb$ et varie entre 0,02% et 0,04% pour le rapport $^{208}Pb/^{206}Pb$. Pour les rapports qui contiennent l'isotope $^{204}Pb$, le pourcentage d'erreur varie entre 0,05 et 0,08%.

En fonction des résultats fournis par les analyses des 20 échantillons (fig. 2), ces derniers se divisent en deux groupes bien distincts, bien qu'il soient proches l'un de l'autre : le premier en comprend 17, le second trois. Dans ce tableau, les résultats des analyses effectuées dans le laboratoire d'Ispra sont donnés sous la forme des rapports isotopiques du plomb.

Dans la figure 3, le diagramme présente les rapports isotopiques $^{206}Pb/^{204}Pb$ *versus* $^{207}Pb/^{204}Pb$ des échantillons de Comacchio et de nombreux minerais provenant de diverses mines d'Espagne et du monde méditerranéen, analysés soit à Ispra soit dans d'autres laboratoires. Ces échantillons couvrent un champ très vaste qui va de 15,5 à 16,1 pour le rapport $^{207}Pb/^{204}Pb$ et de 18 à 18,9 pour le rapport $^{206}Pb/^{204}Pb$. Les droites représentent les isochrones ; comme elles sont tracées d'après le modèle à une seule phase de formation (« single stage model »), elles ne doivent servir que comme références relatives et non point absolues. La figure 4 représente un détail de la précédente, afin de situer plus clairement les échantillons de Comacchio par rapport aux échantillons de minerais les plus proches.

Dans les deux figures 5 et 6, les données présentées sur les diagrammes antérieurs sont reportées d'abord en totalité (fig. 5), puis en détail (fig. 6), sur le diagramme de précision $^{207}Pb/^{206}Pb$ *versus* $^{208}Pb/^{206}Pb$. On peut remarquer que quelques districts miniers peuvent être éliminés comme sources possibles du plomb des lingots de Comacchio. Certains, parce qu'ils sont trop éloignés de ces derniers : ce sont le Laurion, le sud de la France, divers sites de la péninsule Ibérique et même le Pays Basque (principalement les mines d'Arditurri) ; d'autres, parce ce sont essentiellement des districts producteurs de cuivre : Chypre, Sardaigne (mine de Castello di Bonvei : échantillon de malachite). Quant à la Toscane, des preuves certaines d'exploitation existent pour le Moyen Age, mais pas pour l'Antiquité. En revanche, restent en course les mines du sud-est de l'Espagne (fig. 7), où l'on remarquera, à côté du groupe

---

[11] L'échantillonnage a été effectué le 4 avril 2005, au *Museo de la Nave Romana di Comacchio*, par P. Quarati, suite à une Convention signée entre le *Politecnico di Torino* et la *Soprintendenza per i Beni Archeologici dell' Emilia Romagna*. Nous remercions vivement Fede Berti, directrice du *Museo Archeologico Nazionale di Ferrara*, pour son efficace collaboration et sa disponibilité à l'occasion de cette opération.

[12] Les données utilisées pour l'établissement des diagrammes proviennent d'une part des travaux en cours (analyses P.R. Trincherini, JRS, Environment Institute, CCR, Ispra, Varese ; entre autres, celles de la figure 2 « Lingots de Comacchio »), d'autre part de la littérature : Arribas & Tosdal 1994 ; Boni & Koeppel 1985 ; Brévart *et al.* 1982 ; Empresa Adaro s.d. ; Gale *et al.* 1997 ; Graeser & Friedrich 1970 ; Hunt Ortiz 2003 ; Le Guen *et al* ; 1991 ; Ludwig *et al.* 1989 ; Marcoux *et al.* 1992 ; Nesta 2000 ; Oh *et al.* 1989 ; Pernicka *et al.* 1990 ; Santos Zalduegui *et al.* 2004 ; Sinclair *et al.* 1993 ; Stos-Gale *et al.* 1995 ; Trincherini *et al.* 2001 ; Velasco *et al.* 1996 ; Zuffardi 1965.

Les appellations « diagramme de l'âge » et « diagramme de la précision » viennent de Guénette-Beck 2005.

principal des échantillons de Comacchio constitué par 17 lingots, la formation d'un sous-groupe de trois lingots, proche à la fois d'échantillons du Cabo de Gata et d'un échantillon de minerai de Mazarrón.

Dans la figure 8, sont indiquées avec précision les mines d'où proviennent les échantillons du sud-est espagnol proches du groupe principal des lingots de Comacchio. Les mines du Cabo de Gata (La Paniza, San José, Monsu) doivent être éliminées, car elles n'ont pas été exploitées dans l'Antiquité[13]. Restent d'une part les autres mines d'Almería représentées ici, à savoir essentiellement celles de la Sierra Almagrera (Filon Jaroso, Ramo de Flores, Los Hermanos, Guzmana), d'autre part celles du district Cartagena (Pozo del Águila, Gloria, Confianza, Secretario, Mercurio, Cabezo Rajado)-Mazarrón (Cerro San Cristobal, Piedra Amarilla)[14]. À ce niveau, il est difficile d'éliminer soit l'un, soit l'autre de ces deux districts comme source du plomb des lingots de Comacchio.

Enfin, il nous a paru intéressant de représenter sur un diagramme $^{207}Pb/^{206}Pb$ *versus* $^{208}Pb/^{206}Pb$, et sous deux formes différentes (fig. 9 : points ; fig. 10 : nuages) les principaux districts espagnols producteurs de plomb dans l'Antiquité romaine. On y voit par exemple qu'en aucune manière les mines de La Serena (qui avaient été considérées par certains comme une source possible du plomb des lingots de Comacchio) ne peuvent être à l'origine de ces derniers, pas plus que celles de Linares ou des Pedroches-Alcudia. Les lingots de Comacchio, quant à eux, se divisent clairement en deux groupes. Alors que 17 constituent le groupe principal et proviennent, avec une forte probabilité, des mines de Cartagena-Mazarrón ou de celles de la Sierra Almagrera, les trois autres forment un groupe à part, compatible avec Cartagena-Mazarrón.

On peut donc conclure que, du point de vue des isotopes du plomb, tous les lingots de Comacchio sont compatibles avec les mines de Cartagena-Mazzarón et avec celles de la Sierra Almagrera. Mais le plomb dont ils sont faits provient de deux gisements différents, dont un a peut-être été exploité plus intensément que l'autre et qui présentent des caractères géologiques différents. En particulier, les échantillons du groupe des trois lingots contiennent une plus grande quantité de plomb 206 et proviennent de sources plus riches (environ 0,5%) en Uranium 238.

+
+ +

2 – Les timbres

Du coup disparaît le lien supposé des lingots de Comacchio avec les mines de plomb-argent de la Serena (province de Badajoz). *Exit* également le caractère militaire de leur exploitation à l'époque augustéenne, qui découlait à la fois de la fonction défensive que M.P. García Bellido attribuait à ces constructions antiques de cette région appelées par nos collègues espagnols « recintos-torres» et par Pierre Moret « maisons fortes »[15], et du rôle actif qu'elle faisait jouer dans les travaux miniers à d'hypothétiques détachements des légions I, IV et X, qui, en 19 av. J.-C., auraient été placés sous le commandement d'Agrippa.

---

[13] Depuis 1995, date à laquelle elles sont apparues massivement dans la littérature concernant les isotopes du plomb (Stos-Gale et al. 1995), les mines du Cabo de Gata (Almería) posent un problème qu'il n'est pas possible de traiter ici. Il le sera ailleurs. Disons pour l'instant qu'il n'y a pas la moindre trace de présence romaine au voisinage des mines de plomb de ce district, composé de petites mines qui n'ont été exploitées, semble-t-il, qu'au XIX$^e$ et au XX$^e$ siècle.

[14] Pour les mines de la Sierra Almagrera et celles de Cartagena-Mazarrón dans l'Antiquité, voir Domergue 1987, p. 5-8 et 358-390.

[15] García Bellido 1994-1995, p. 206-213. En fait, le caractère militaire de ces constructions n'est pas si évident. Il est nié, avec de bons arguments, en particulier par P. Moret (1995) et C. Fabião (2002). Pour ces auteurs, il s'agit plus vraisemblablement d'un type d'architecture rurale, d'inspiration à la fois italique et hellénistico-romaine (P. Moret) qui s'est diffusé en Hispanie à partir de l'époque césarienne (Alentejo portugais) et sous le Haut-Empire (La Serena) ; il serait la marque d'un modèle d'occupation du sol, au même titre que la *villa*, mais dans des zones différentes et peut-être suite à des choix politiques différents (C. Fabião).

Il faut donc, me semble-t-il, renoncer à l'interprétation « légionnaire » des lingots de Comacchio, et revenir à une lecture « civile » des timbres qu'ils portent. Rappelons d'abord que, s'agissant de lingots de plomb, si les estampilles moulées dans les cartouches dorsaux désignent les producteurs, les timbres imprimés après démoulage sur la surface du lingot sont ceux des acheteurs et possesseurs successifs de ces marchandises[16].

Il n'y a pas ici d'estampille moulée renfermant le nom du producteur, mais seulement, dans le cartouche dorsal des séries IV et V, un caducée moulé (fig. 11). En revanche, les timbres imprimés à froid sont nombreux.

2.1. La revue des timbres

Avant d'entreprendre l'examen de cette série de timbres, il convient de rappeler que, pour l'instant, deux districts miniers du sud-est, Cartagena-Mazarrón et la Sierra Almagrera, l'un et l'autre exploités au cours de l'Antiquité romaine, sont en compétition pour savoir lequel des deux est celui d'où proviennent les lingots, puisque les analyses isotopiques n'ont pu permettre de le déterminer. Or l'étude épigraphique peut aider à cette détermination. On peut en tout cas émettre l'hypothèse selon laquelle l'épigraphie des lingots doit refléter celle de la région de production. Vérifions-le donc en prenant comme second terme de comparaison l'épigraphie de *Carthago Noua*. Comme elle est particulièrement riche en gentilices liés à la production de plomb, le test a des chances de réussir, que la réponse soit positive ou négative. Simplement, dans le second cas, il faudra se tourner vers la Sierra Almagrera.

Il y a dix timbres au total (fig. 12). Certains (n° 1, 2, 3, 4, 5 et 6) sont parfaitement lisibles, et, même si leur signification pose des problèmes, ils ne surprennent pas quelqu'un qui soit peu ou prou versé dans l'épigraphie latine : ils paraissent désigner des individus, certains plus clairement que d'autres. Il n'en va pas de même pour les quatre suivants, qu'on ne sait par quel bout prendre : ils sont un peu comme des rébus.

Commençons donc par la série « des rébus » et, dans cette dernière, par le n° 8, car il va nous donner une clé pour en interpréter quelques autres. De toute évidence, ce timbre est composé d'une suite de lettres, mais elle est peu compréhensible si on cherche à la lire telle qu'elle se présente. Considérons-la plutôt à l'envers : on lit alors clairement FVRI, avec ligature du F, du V et du R. Les Furii sont connus à *Carthago Noua*, où ils sont attestés épigraphiquement[17] et où ils ont exploité des mines de plomb-argent à la fin de la République, comme en témoigne un lingot de plomb estampillé au nom de A. et P. Furii, tous deux affranchis de C., L., et P. Furii [18]. Dans ce cas, le test a fort bien fonctionné ; on remarquera même que, sur l'estampille du lingot, on retrouve la même ligature : FVR.

Dans cette même ligne, le timbre 10 peut fort bien être considéré comme un groupe de trois lettres ligaturées dans le sens de la hauteur : VTI. Ici encore, il s'agit d'une famille bien connue de producteurs de plomb de *Carthago Noua*, dont le nom est attesté sur des lingots trouvés en divers points de la Méditerranée : port de Carthagène, épave de la Madrague de Giens (environ 65 avant J.-C.), etc.[19]

---

Dans des zones minières comme la Serena ou l'Alentejo, ces constructions n'ont pas de lien direct avec les mines, dont elles sont souvent très éloignées. Quant à leur fonction, P. Moret (1995, p. 546), pour qui ces « maisons fortes » sont des fermes, a suggéré qu'il pouvait s'agir « de petites exploitations agricoles capables d'approvisionner les campements et les villages miniers en grain, en viande et en laine ».

[16] Domergue 1998, p. 203.
[17] *CIL* II 3467 et 3468 = Abascal & Ramallo 1997, n° 139 et 140.
[18] Trouvé dans le golfe de Fos/Mer (Bouches-du-Rhône, France) (Liou 1992).
[19] Domergue 1990, 256, n° 1052 à 1054.

Jetons un regard semblable sur le timbre 9. Sous la figure géométrique apparente (les deux petits côtés et les diagonales d'un rectangle), nous croyons pouvoir déceler une ligature de quatre lettres, IVNI : le petit côté gauche constitue à la fois la haste du I, la branche gauche du V et la barre gauche du N, la diagonale qui va de gauche à droite et de bas en haut est la deuxième branche du V ; la deuxième diagonale est la barre oblique du N, dont la barre verticale droite est constituée par le petit côté droit, qui sert également de I final. Cette signature a des chances de renvoyer à la famille des Iunii Paeti, qui, au Ier siècle avant J.-C., produisait du plomb à *Carthago Noua*, comme le prouve la découverte, au Magdalensberg, d'un lingot estampillé au nom de C. Iunius Paetus[20], tandis que, précisément à l'extrême fin du Ier siècle avant J.-C., un autre de ses membres, L. Iunius Paetus, peut-être un neveu du précédent, dédie, dans le théâtre de *Carthago Noua*, deux autels en marbre l'un à la Fortune, l'autre à Caius Caesar[21].

La ressemblance du timbre 9 et du timbre 9a[22], le fait que l'on ne les trouve pas ensemble (fig. 13), enfin le fait que la seule occurrence du n° 9 est, comme la plupart de celles du n° 9a, associée au timbre 3 et au timbre 8 nous amènent à considérer que le second doit être une simple variante du premier : sur la matrice du timbre 9a, les deux hastes verticales qui unissent les extrémités des deux diagonales auront été omises ou insuffisamment tracées, voire mal comprises par l'auteur de cette matrice du timbre, qui les aura supprimées.

**Ici fig. 13 (tableau)**

Le timbre 7 est constitué par une ligature de trois ou de quatre lettres, dont la première est un P (*retro*). Il s'agit d'abord de savoir comment on interprète la haste droite médiane : s'agit-il d'un I longa ou bien l'allongement est-il simplement dû au fait qu'elle sert de haste verticale aux trois lettres P, L et R ? Ce peut être dans les deux cas l'abréviation d'une dénomination sous la forme de tria nomina : P(ublius) Li(…) R(…) ou P(ublius) L(…) R(…) ; pour le gentilice, l'épigraphie de *Carthago Noua* offre un choix assez étendu, de Labicius à Lumnesius[23] en passant par Laelius, Laetilius, Licinius, Lucretius, des nomina dont certains (Laetilius, Lucretius) sont attestés parmi les producteurs de lingots de *Carthago Noua* [24]. Mais, vu la composition des cachets 8, 9 et 10, où, selon notre lecture, le gentilice apparaît sans prénom, nous préférons lire PLR et développer ces trois lettres en Pl(anii) R(ussini), autrement dit le nom et le surnom d'une des grandes familles productrices de plomb à *Carthago Noua* au Ier siècle avant notre ère[25].

Si l'on se rend aux arguments qui précèdent, quatre des dix timbres recensés sur les lingots de Comacchio admettent une lecture et un développement qui tendent à privilégier les mines de *Carthago Noua* comme district d'origine des lingots de Comacchio, au moins de ceux qui portent l'un des quatre timbres examinés et qui appartiennent aux séries I, II, II et IV. Certes, prises isolément, ces lectures n'ont pas une égale force de persuasion, mais celle du premier timbre paraît incontournable et entraîne logiquement l'adhésion aux trois autres, donnant à ce bloc épigraphique que constituent les quatre timbres considérés une cohérence satisfaisante, par une espèce d'effet « boule de neige ». Remarquons cependant que les quatre timbres qui renvoient à *Carthago Noua* concernent les séries de lingots I, II, III et IV ; la série

---

[20] Piccottini *et al.* 2003 ; Domergue & Piccottini 2004.
[21] Abascal & Ramallo 1997, n° 12 et 13 ; Ramallo 2003.
[22] F. Berti (1987, p. 132) rapproche ce timbre d'un de ceux qui marquent les lingots d'étain de l'épave *Bagaud 2* (fin du IIe-début du Ier siècle avant J.-C.). Effectivement la ressemblance est frappante, mais les dates sont différentes Ici, en revanche, le rapprochement avec le timbre n° 9, qui fait partie du même contexte, me paraît préférable.
[23] Abascal & Ramallo 1997, p. 533-534.
[24] Domergue 1990, 254-257, n° 1022, 1015, 1046
[25] Domergue 1990, p. 255, n° 1028 à 1031

V est à l'écart, comme les analyses de trois lingots de cette série étaient, elles aussi, à l'écart des 17 autres, et plus proches d'un des champs isotopiques de Mazarrón que de celui des mines de *Carthago Noua*. Mais qu'en est-il des 6 timbres restants ?

Nous mettons à part le timbre n° 1, AGRIP, qui apparaît sur 92 lingots : il peut désigner un Agrippa quelconque et l'on a effectivement souligné la relative fréquence de ce cognomen[26], mais, selon l'interprétation que nous en avions donnée en 1987[27] et qui s'accompagnait d'une argumentation épigraphique nourrie, il peut s'agir aussi de M. Agrippa, le gendre d'Auguste. On ne peut d'autre part ignorer aujourd'hui la convergence de trois arguments : le matériel de l'épave confirme bien une date de naufrage vers la fin du Ier siècle av. J.-C., la signature isotopique du plomb indique comme provenance principale les mines de *Carthago Noua*, enfin Agrippa a été patron de la colonie de *Carthago Noua*[28] ; à ce titre, il a des rapports privilégiés avec cette cité, il a donc pu vouloir se procurer du plomb auprès d'elle, étant donné la réputation qu'avaient naguère les mines voisines. Nous reviendrons plus loin sur cette question.

L. CAE. BAT : ce timbre est porté par 96 lingots. Il accompagne donc couramment le précédent. Ils sont présents, soit les deux à la fois, soit l'un, soit l'autre sur tous les lingots sauf deux[29]: les deux timbres paraissent donc liés. On avait proposé de voir dans celui-ci la marque d'un familier d'Agrippa et de développer CAE. en Cae(cilius), le personnage ainsi désigné pouvait être un affranchi de la première épouse d'Agrippa, Caecilia Attica, ou du père de cette dernière, T. Pomponius Atticus, qui, après avoir été adopté par son oncle, était devenu Q. Caecilius Pomponianus Atticus[30]. On attendrait cependant pour un tel personnage un cognomen latin, mais on n'en connaît guère qui commencent par ces trois lettres, alors qu'elles s'accordent fort bien avec un nom originaire de la péninsule Ibérique, *Bat-ia, -icus, -ialus*, des noms peut-être lusitaniens[31]. Si par ailleurs, on envisage qu'il puisse s'agir de quelqu'un de *Carthago Noua*, deux gentilices commençant par CAE sont attestés dans l'épigraphie de cette cité, où ils sont ceux de personnages qui ont fait partie de l'élite municipale à l'époque augustéenne : le premier est Caesius[32]. Un Caesius de *Carthago Noua* en effet a son nom gravé sur une base honorifique datée de cette période[33]. Le second, Caedius, est celui d'une grande famille de *Carthago Noua*, dont un membre fut duumvir quinquennalis et magistrat monétaire entre 42/41 av. J.-C.[34]. On peut donc parfaitement imaginer, que notre L. Cae(…) Bat(…) désigne un L. Cae(sius) ou Cae(dius) Bat(ialus ?), sans doute un affranchi de l'une ou de l'autre gens ; enfin, que son cognomen puisse être d'origine lusitanienne n'est pas une gêne : dans la péninsule Ibérique, les gens circulent, comme le montre le main-d'œuvre employée à la mine de La Loba, à la fin du IIe-début du

---

[26] Bruun 1991, p. 362, n. 12.
[27] Domergue 1987 a, p. 118-122.
[28] Abascal & Ramallo 1997, n° 42. On ignore la date exacte de la fondation de la *Colonia Iulia C.N.*, mais, en tout état de cause, elle est antérieure à la mort de César (43 avant J.-C.) (Abascal & Ramallo 1997, p. 14-15). Agrippa est devenu *patronus* de *Carthago Noua* entre 19/18 et 12 avant J.-C..
[29] Ces deux lingots (1 et 88) n'ont pas été nettoyés ; les concrétions doivent donc dissimuler les timbres qu'ils portent.
[30] Domergue 1987 a, p. 122-123.
[31] Albertos Firmat 1965, p. 51 ; ILER, 1231 et 5501. *Bat(-avus)*, ethnique germain, n'est attesté que dans le Norique et à Trèves (Kajanto 1965, 50 et 201) et doit être plus tardif.
[32] C'est aussi celui qu'avait retenu M.P. García Bellido pour développer CAE, car, bien attesté en Lusitanie, où les Caesii sont sans doute des descendants de clients de L. Caesius, gouverneur de l'Ultérieure en 104 av. J.-C. (García Bellido 1998a, p. 19-20), et dans l'ouest de la Bétique, il pouvait désigner un personnage en relation avec la région minière d'où venaient, selon cet auteur, les lingots de Comacchio : La Serena (García Bellido 1998a, p. 37). Pour notre part, nous nous engageons dans une direction toute différente.
[33] Abascal & Ramallo 1997, n° 58.
[34] Selon Llorens 1994, p. 41 et 144.

Ier siècle avant J.-C.[35]. En tout cas, qu'il soit originaire d'Italie et s'appelle Caecilius, ou, comme nous le préférons aujourd'hui, qu'il soit de *Carthago Noua* et que son nom soit Caesius ou Caedius, ce personnage est le fidèle doublon d'Agrippa, vraisemblablement son associé.

Le timbre 3, GEME, avec ligature du M et du E[36], est relativement fréquent, puisqu'on le trouve sur vingt lingots (2, 7, 10, 11, 13, 16, 18, 22, 23, 26, 30, 32, 33, 34, 37, 38, 51, 52, 54, 55). Aucun développement ne s'impose, le plus probable est qu'il s'agisse d'un nom, *Geme(-llus, -llinus, -llianus ?)*[37], nom d'esclave ou surnom d'affranchi ou d'ingénu. Rien dans l'épigraphie de *Carthago Noua* ne permet de compléter l'abréviation.

Le timbre 5, C. MATI est beaucoup moins représenté : seulement quatre occurrences, une fois en compagnie du timbre n° 9a (25)[38], trois fois en compagnie du timbre 4, MAC (3, 4, 17). Il désigne un certain Gaius Matius, au génitif[39]. À l'heure actuelle, le gentilice *Matius* n'est pas attesté à *Carthago Noua*. L'hypothèse que l'un de nous avait émise en 1987 – voir dans C. Matius sinon le chevalier ami de César, puis d'Auguste[40], mais plutôt un de ses affranchis – ne s'impose pas dès lors qu'on ne cherche pas à tout interpréter à travers le prisme romain. Notre C. Matius paraît donc être inconnu.

Pour le timbre 4, MAC, il n'y a guère de développement possible en dehors d'un nom d'esclave ou d'un cognomen d'affranchi : Mac(edo), Mac(er), Mac(rinus), etc. Il apparaît sur onze lingots (3, 4, 11, 14, 17, 20, 38, 39, 41, 49, 53), et il accompagne C. MATI sur trois d'entre eux (3, 4, 11). Aussi, et vu la similitude des dimensions de deux timbres 5 et 4, nous rapprocherions volontiers ces derniers, comme déjà en 1987, pour obtenir une dénomination plus complète : C. Matius Mac(…). Dans les cas où MAC est seul, il doit suffire à représenter C. Matius.

Le timbre 6, MAT, est rare, puisqu'on ne le trouve que sur trois lingots (12, 15, 29), où il est seul[41]. Il est difficile de le développer. Une solution simple : ce pourrait être un substitut du timbre 5 et constituer une troisième marque de C. Matius.

Quatre timbres (n° 7, 8, 9/9a, 10) paraissent donc liés directement à l'épigraphie de *Carthago Noua*, ce qui permet de rattacher aux mines de ce district au moins les lingots ainsi marqués des séries I, II, III et IV, et, très vraisemblablement, tous ceux de ces quatre séries Ce n'est pas le cas des autres timbres, sauf peut-être du n° 2 (et encore le lien évoqué est-il seulement hypothétique), le n° 1 étant à part.

2_2. Observations sur les timbres et sur leur disposition

Nous allons d'abord faire de rapides remarques sur la distribution des timbres (dont on trouvera le détail dans le tableau de la figure 13), étant entendu que quelques lingots portent encore des concrétions qui peuvent en dissimuler certains, mais, sauf peut-être dans le cas des timbres peu représentés, une occurrence nouvelle ne devrait pas changer l'interprétation que l'on pourra en faire ; ensuite considérer leur emplacement ; enfin envisager les cas de superposition de deux, voire trois timbres : ces trois types d'observations peuvent en effet aider à comprendre le rôle respectif des personnages représentés par les timbres.

---

[35] Domergue & Sillières 2002, p. 394-395.
[36] GEME et non GEM (cf. fig. 3) : les barres horizontales du E, quoique courtes, sont bien visibles sur les exemplaires les plus complets.
[37] Gemellus : trois exemples d'époque républicaine et de nombreux autres dans l'ensemble du *CIL*, (Kajanto 1965, p. 295).
[38] Les chiffres entre parenthèses renvoient à l'inventaire du tableau de la figure 13.
[39] Je ne pense pas que, vu la date, cette désinence en –*i*, comme d'ailleurs toutes celles qui figurent dans nos timbres, soit celle de nominatifs singuliers archaïques (voir infra, note 45).
[40] Domergue 1987 a, p. 124-125.
[41] Ici et dans les lignes qui suivent, l'adjectif « seul » s'entend à l'exclusion des timbres 1 et 2 qui sont présents sur la quasi-totalité des lingots.

2.2.1 – La distribution des timbres (fig. 13) :
*a* - sur deux lingots (1 et 88), aucun timbre n'est visible, mais ils sont couverts d'une couche de concrétions ; nous ne les prenons pas en considération ;
*b* - timbres 1 (AGRIP) et 2 (L.CAE.BAT) : au moins l'un des deux timbres est présent sur chaque lingot (le premier sur 92 lingots, le second sur 96) ; ils sont souvent présents tous les deux ensemble et sont attestés dans les cinq séries ;
*c* - les timbres « *Carthago Noua* » : les n° 8 (FVRI), 9 et 9a (IVNI), 10 (VTI) apparaissent sur des lingots des séries I et II : les n° 8 et 9/9a associés sur neuf lingots (10, 13, 16, 18, 23, 26, 28, 32, 33), le 9a seul sur un seul autre (21), on peut donc considérer qu'ils vont ensemble. Le timbre 10 est présent sur un (9) et sans doute deux lingots (19). Le timbre 7 (PL[ani] R[vssini]) est visible sur la quasi-totalité de la série IV (à l'exception du 88) ; il apparaît aussi sur quelques lingots dans les séries I (5, 8), II (27, 35) et III (56) ;
*d* - les timbres « autres » : les n° 5 (C. MATI), 6 (MAT), 3 (GEME), 4 (MAC) sont marqués exclusivement sur les lingots des séries 1, 2 et 3. Ils sont parfois seuls : MAC (14, 20, 39, 41, 49, 53), MAT (n° 12, 15, 29), GEME (2, 7, 22, 30, 34, 37, 51, 52, 54, 55). Ils peuvent apparaître avec un autre de ce genre : C. MATI + MAC (3, 4, 17), GEME + MAC (11, 38). Les timbres 5 et 3 peuvent accompagner des timbres « *Carthago Noua* » : on rencontre C. MATI avec le timbre 9a (IVNI) sur un seul lingot (25), GEME avec les timbres 8 (FVRI) et 9/9a (IVNI) sur les lingots 10, 13, 16, 18, 23, 26, 32 et 33. Les timbres 4 et 6 ne se trouvent jamais sur les mêmes lingots que les timbres « *Carthago Noua* ».

Ces observations n'ont pas toutes la même importance. On remarquera d'abord l'absence de timbres (les n° 1 et 2 mis à part) sur plusieurs lingots : tous ceux de la série V, et quelques exemplaires des séries I (6), II (24, 31, 36, 40, 42) et III (43 à 48, 50). Est-ce à dire que les « personnages-timbres » 1 et 2[42] sont directement entrés en possession de ces lingots ? C'est peut-être le cas pour ceux de la série V, typologiquement homogène, mais c'est plus difficile pour ceux des séries I à III, qui sont beaucoup plus hétérogènes et dont plusieurs lingots portent des timbres « *Carthago Noua* ». On peut en conclure que tous les lingots n'étaient pas systématiquement marqués par chacun de leurs possesseurs successifs[43].

Considérons ensuite les timbres « *Carthago Noua* » : aucun des trois timbres 8, 9/9a et 10 n'apparaît sur les lingots des séries IV et V : cela signifie que ces lingots ne sont pas passés entre les mains des personnages qui correspondent à ces trois timbres. En revanche, les lingots de la série IV, qui portent tous le timbre 7, ont dû être, à un moment donné, la propriété de ce personnage-timbre. Ce même timbre 7 apparaît, également seul, sur quelques lingots des séries I à III (*c*), qui ne portent aucun des trois autres timbres « *Carthago Noua* : ils lui ont appartenu aussi.

Les timbres « autres » n'apparaissent jamais dans les séries IV et V, cela souligne l'intérêt du personnage-timbre 7. On trouve en revanche les timbres 3, 4, 5 et 6 dans les trois premières séries. L'un d'eux (n° 3) est particulièrement fréquent et accompagne souvent les timbres « *Carthago Noua* », tandis que les n° 4 et 6 n'apparaissent jamais en compagnie de ces derniers.

---

[42] « Personnage-timbre » : nous nous permettons cette expression barbare par commodité, afin d'éviter les circonlocutions du genre « les personnages qui sont représentés par le timbre 3 », etc.
[43] Cette observation n'est pas nouvelle : cf. les lingots des épaves *Cabrera 5* (Colls *et al.* 1986, 61-63), *Sud-Perduto 2* (Bernard & Domergue 1991, 78-79). Mais il vaut la peine de noter qu'elle est valable ici aussi. On peut expliquer ce fait de diverses façons : par exemple, s'agissant d'un tas de lingots, il suffisait que son propriétaire apposât sa marque seulement sur ceux qui se trouvaient sur le dessus.

2.2.2   L'emplacement des timbres (fig. 13)

Dans les séries I à III, deux emplacements sont utilisés : la base et le dos. Dans les séries IV et V, deux également : la base ou un long côté[44]. Il est évident que cela dépendait de la façon dont les lingots étaient disposés, sur le dos ou sur la base, face aux personnages chargés du marquage. Parfois, les lingots ont pu être posés sans ordre et dans les deux sens. En revanche il y a des cas où il y a eu un choix, par exemple lorsqu'un timbre est exclusivement posé sur la base ou sur le dos de tous les lingots - ou, au moins, d'un nombre significatif de lingots - d'une série. Voyons maintenant le détail :

*e*– les timbres 2 (sur 96 lingots au total) et 7 (sur tous les exemplaires de la série IV et quelques exemplaires des séries I à III) sont toujours imprimés sur la base : c'est sûrement un choix : une surface plane se prêtait davantage à cette opération. Mais le fait que le timbre 7 n'ait pas été posé sur un aussi grand nombre de lingots que le 2 suggère que ces deux timbres n'ont pas été imprimés au même moment, sinon pourquoi le marqueur aurait-il négligé de marquer du timbre 7 les lingots de la série V et ceux des séries I à III qui en sont privés?
*f* – Il est vraisemblable que les timbres 1 et 2, qui, comme on l'a vu, constituent un tandem, n'ont pas été imprimés en même temps ; dans le cas contraire en effet, on ne comprendrait pas pourquoi, sur plusieurs lingots, le timbre 1 a été imprimé sur le dos (séries I à III) ou sur un côté (séries IV et V), alors que le 2 est toujours posé dur la base (*e*) : pourquoi aurait-on retourné certains lingots et pas d'autres, alors qu'il était plus facile d'imprimer le timbre sur la surface plane de la base ?
*g* – Les timbres 8 et 9/9a sont toujours imprimés sur le dos, à une seule exception près pour 9a (21). Le timbre 3 (GEME) les accompagne à chaque fois, et il est également imprimé sur le dos, sauf dans le cas du lingot 21, où il n'apparaît pas. Cela peut signaler un rapport plus ou moins étroit, par exemple qu'il était nécessaire que fussent vus d'un seul coup d'œil les timbres 8 et 9/9a avec le timbre 3. En revanche, toutes les autres occurrences de ce timbre 3 figurent sur la base de lingots (2, 34, 37, 38, 51, 52, 54, 55).
*h* – Timbre 3 (GEME) et timbre 1 (AGRIP) : sur les lingots où l'un et l'autre figurent, le premier est majoritairement imprimé sur le dos, le second sur la base. Malgré leur relation (*k*), ils n'ont pas dû être imprimés en même temps
*i* – On ne peut dire grand chose sur les timbres 5 et 6, car il y a trop peu d'occurrences. Ils sont imprimés sur les bases, sauf en un cas (25), où C.MATI a été apposé sur le dos.

2.2.3 – La superposition de timbres (fig. 14)

Nous rassemblons dans le tableau ci-dessous les cas de superposition de timbres :

<span style="color:red">Ici figure 14 (tableau)</span>

On peut tirer de ce tableau les enseignements suivants :
*j* – vu les lignes 1 et 2, les timbres AGRIP et L.CAE.BAT sembleraient avoir été imprimés au cours de la même phase de marquage ; pourtant, il n'en est sans doute rien (*f*), quelle que soit la série considérée. Mais, étant donné leur distribution (*b*), ils paraissent interchangeables : les deux personnages correspondants constituent un couple d'associés ;

---

[44] À l'exception du lingot 65, où le timbre 1 figure sur un petit côté.

*k* – vu les lignes 3 et 4, les timbres AGRIP et GEME pourraient avoir été imprimés au cours de la même phase de marquage. Mais rien n'est moins sûr . En fait, ce qu'exprime cette relation réversible entre les deux timbres, c'est sans doute que tous deux devaient être imprimés sur les lingots concernés, quel que soit le moment où chacun d'eux l'était. Geme(…) doit donc être lié de quelque façon à Agrippa et, par conséquent, au couple d'associés Agrippa et L. Cae(…) Bat(…) ;
*l* – vu les lignes 5 et 6, le timbre « *Carthago Noua* » n° 7 (PLR) est antérieur aux timbres AGRIP et L.CAE.BAT. Mais les considérations sur l'emplacement respectif des timbres 2 et 7 d'une part (*e*), 1 et 2 de l'autre (*b*) semblent indiquer que l'impression de chacun de ces trois timbres a dû se passer à des moments différents.
*m* – vu les lignes 7 et 8, les timbres MAC (n° 4) et MAT (n° 6) ont été été imprimés après ceux des associés Agrippa et L. Cae(…) Bat(…). Cette observation implique que ces deux personnages-timbres sont intervenus les derniers sur les deux lingots en question (15, 41) et pourrait suggérer qu'ils ont été les derniers possesseurs des lingots.
*n* – seul de la série de timbres autres, C.MATI (n° 5) échappe à toute superposition. Il en va de même pour les timbres « *Carthago Noua* » 8, 9/9a, 10, mais la situation de ces derniers doit être différente, comme nous le verrons plus loin. Quant au personnage désigné par le timbre 5, nous avons plus haut émis l'hypothèse qu'il pourrait aussi être désigné par les timbres n° 4 (cognomen) et 6.

Les observations qui précèdent sont de portée variable. Les plus intéressantes portent sur la distribution et la superposition des timbres, en particulier pour ce qui concerne l'ordre dans lequel ils ont été imprimés et les rapports qui unissent certains d'entre eux. Les remarques faites à propos de l'emplacement des timbres sont d'un intérêt moindre, mais certaines apportent parfois des compléments aux précédentes. En tout cas, le schéma auquel nous sommes conduits paraît être le suivant : les personnages-timbres n° 7, 8, 9/9a et 10 (Plani Russini, Fvrii, Iunii et Vtii[45]) ont chacun fourni, et de manière inégale, des lingots de plomb (au total 102) au couple d'associés Agrippa et L.Cae(…) Bat(…) (timbres 1 et 2) ; certains l'ont fait directement (n° 7 et sans doute 10[46]), les deux autres, qui, par ailleurs semblent travailler ensemble (sont-ils associés ?) (*c*, *g*), par l'intermédiaire du personnage-timbre 3 (GEME), qui est sans doute un agent d'Agrippa et de L. Cae(…) Bat(…), pour lesquels il a joué le rôle de « rabatteur ».

Les timbres 4 et 6, qui nous paraissent renvoyer au n° 5, représenteraient, avec ce dernier, le ou les derniers possesseurs des lingots, à qui Agrippa et L.Cae(…) Bat(…) avaient fini par revendre tout le stock[47].

---

[45] Une précision s'impose ici, : les timbres 7, 8, 9 et 10 sont-ils au nominatif singulier, au génitif singulier ou au nominatif pluriel ? Nous exclurions volontiers les deux premières hypothèses : d'une part un nominatif singulier archaïque en –*i*, bien attesté à *Carthago Noua* un bon siècle plus tôt (González Fernández 1995), est peu probable ici, vu la date ; de plus, aucun de ces *nomina* n'est individualisé par un prénom (il en va différemment dans le cas de C. Mati). Le *nomen* doit donc être considéré comme global : il désignerait ceux des Furii, des Iunii, des Planii Russini, des Vtii qui se sont engagés dans le négoce du plomb, comme on connaissait auparavant, en tant que producteurs de plomb, des Aquinii, Lucretii, Planii, Pontilieni et autres Roscii, dont les noms, parfois accompagnés des initiales de plusieurs prénoms, se lisaient sur les estampilles moulées de nombre de lingots de *Carthago Noua* « de la belle époque » (Domergue 1990, p. 254-257).
[46] Pourquoi « sans doute » ? Seuls deux lingots portent le timbre VTI, et l'on sait que le marquage des timbres n'est pas systématique…
[47] Ces trois timbres posent plusieurs problèmes :
1 - Si l'on considère le tableau de la figure 13, la première idée qui vient à l'esprit est que ces timbres jouent dans les séries I, II et II un rôle analogue à celui des timbres « *Carthago Noua* », autrement dit que C. Matius pourrait être, au même titre que les Aquinii, Lucretii, Planii, Pontilieni et Roscii, un négociant en plomb de cette cité. Mais d'une part, à la différence des autres, ce gentilice n'est attesté ni dans l'épigraphie de *Carthago Noua* ni dans celle des producteurs de plomb de cette cité. D'autre part, alors que le timbre des Planii Russini a été

+
+   +

3 – Les avatars des lingots de Comacchio : des mines de *Carthago Noua* à l'Adriatique nord.

Le bateau de Comacchio a fait naufrage à la fin du Ier siècle avant J.-C., plus précisément dans l'avant-dernière décennie, entre les années 19 et 12, si nous considérons que l'Agrippa dont le timbre est visible sur les lingots de plomb de Comacchio est bien M. Vipsanius Agrippa, le gendre d'Auguste. Nous avons rappelé plus haut les raisons qui rendent possible cette identification, que nous maintenons. Agrippa, soit qu'il se soit lancé en tant que particulier dans le commerce du plomb, soit qu'il ait besoin de ce métal pour les opérations d'urbanisme qu'il conduit officiellement à Rome[48], cherche à s'en procurer. Il connaît la réputation des mines de *Carthago Noua*, une ville dont il est le *patronus*, peut-être depuis 19/18 avant J.-C., et *duumvir quinquennalis* plus récemment sans doute[49]. Dans cette opération, il semble s'être associé à un personnage, qui, s'il est originaire de *Carthago Noua*, pourrait s'appeler L. Caedius ou Caesius Bat(ia, -icus ou -ialus?) et qui pouvait, plus que tout autre, lui faciliter les choses. Car les mines de *Carthago Noua* ne sont plus ce qu'elles étaient. C'en est fini de leur grande prospérité qui fut certaine dans la deuxième moitié du IIe siècle avant J.-C. et dans la première moitié du Ier: c'était l'époque où elles étaient propriété de l'État romain et où de grandes familles minières – les Atellii, Aquinii, Iunii Paeti, Lucretii, Planii Russini, Pontilieni et autres Vtii - régnaient sur la production du plomb de *Carthago Noua* et couvraient le monde méditerranéen de leurs lingots standardisés, aux magnifiques estampilles moulées dans des cartouches dorsaux, parfois accompagnées de ces symboles protecteurs de la navigation que sont l'ancre, le dauphin, le gouvernail ou le caducée. Désormais, ces lingots ont disparu de la circulation commerciale.

Une telle situation semble traduire une soudaine baisse d'activité des mines de *Carthago Noua*, dont il est difficile de trouver les raisons. Faut-il invoquer l'épuisement des gisements ? Il est peu probable, comme le montrent l'exploitation des XIXe-XXe siècles , les vestiges archéologiques de travaux, principalement métallurgiques, situés au pied nord de la Sierra et datés du Ier siècle de notre ère[50], et enfin les lingots de Comacchio eux-mêmes. Les difficultés techniques, l'évacuation des eaux souterraines, les trop grandes profondeurs

---

apposé sur les lingots antérieurement à ceux d'Agrippa et de L. Cae(…) Bat(…) et qu'il est loisible de penser qu'il en va de même pour les autres timbres « *Carthago Noua* » 8, 9 et 10, deux des trois timbres rattachés à C. Matius sont postérieurs à celui d'Agrippa, et donc à celui de L. Cae(…) Bat(…). C. Matius ne saurait donc être un négociant en plomb de *Carthago Noua*, fournisseur de base de ces derniers.
2 - Autre hypothèse : ne pourrait-on pas considérer C. Matius comme un autre Gemellus, c'est-à-dire comme un agent des deux associés ? En réalité, la situation de ces deux personnages n'est pas la même, compte tenu des cas de superposition de leurs timbres (fig. 14). Ce qui nous a permis d'avancer cette hypothèse dans le cas de Gemellus, c'est la relation réversible entre le timbre GEME et et le timbre AGRIP (cf. *k*), mais il n'en va pas de même des rapports de MAC et de MAT avec AGRIP, puisque, si, dans deux cas (fig. 14, ligne 7 et 8) le timbre AGRIP a été imprimé antérieurement à MAC ou à MAT, l'inverse n'est pas attesté.
Ainsi, à nous en tenir aux faits tels qu'ils sont résumés dans le tableau de la figure 14, la seule solution que l'on puisse envisager, c'est que C. Matius soit bien le dernier propriétaire des lingots.
3 - On pourrait objecter que les timbres de ce personnage n'apparaissent que sur tout petit nombre de lingots (15 au total, dans les séries I à III) et jamais sur ceux des séries IV et V (46 lingots à elles deux), mais tout dépend de la façon dont se présentait le lot au moment du marquage : s'ils formaient un tas, seuls ceux qui étaient dessus ont reçu un ou deux de ces timbres, et cela devait suffire (cf. aussi note 42, *supra*).
[48] La chose ne paraît pas impossible à Bruun 1991, p. 362-363.
[49] Llorens 1994, p. 59-61.
[50] Antolinos 1998 (2005), p. 585-587 ; Orejas & Antolinos 1999, II 1a.

atteintes ? Peut-être ces facteurs ont-ils joué un rôle ; ou encore des événements extérieurs : les remous de la guerre civile ont pu déstabiliser le régime d'exploitation et l'engagement des entreprises italiennes.

En tout cas, cette situation coïncide avec le changement de statut de ces mines dont Strabon (3, 2, 10) se fait l'écho. De cette période en effet – fin de l'époque césarienne, début de la période augustéenne – paraît dater le passage des mines de plomb-argent de *Carthago Noua* du domaine public au domaine privé. Les deux séries d'événements sont-elles liées ? Ce n'est pas impossible : à partir du moment où ces mines – pour quelque raison que ce fût[51] - ne lui rapportaient plus, l'État romain s'en est défait. Cela a pu avoir lieu par exemple à l'occasion de la promotion de *Carthago Noua* au statut colonial[52], sans doute avant 43 avant J.-C.. Une partie des mines du domaine public (*ager publicus*) a pu alors être attribuée à la nouvelle colonie, qui en a exploité certaines pour son propre compte (comme le montreraient les deux lingots estampillés *CARTHAGO NOVA* trouvés à Riotinto[53]) et a pu louer les autres à des entrepreneurs privés dans le cadre d'un contrat de *locatio-conductio*. Mais une partie seulement, car les termes employés par Strabon sont très nets : il parle d'*idiôtikai ktèseis*, ce qui implique bien que désormais, des particuliers sont véritablement propriétaires de ces mines et non simples locataires.

La réalité archéologique ne reflète guère cette situation, à l'exception des lingots de Riotinto déjà cités. Qu'en est-il de ceux de Comacchio, qui correspondent bien à cette nouvelle période ? Ils ne portent pas d'estampille moulée désignant nommément un producteur. Seul, le caducée des estampille des séries IV et V (fig. 11) semble signifier quelque chose : est-il la marque anonyme d'un entrepreneur particulier ? Ou bien la colonie de *Carthago Noua* aurait-elle adopté ce symbole, rendu célèbre par des producteurs antérieurs (par exemple, C. Nonius Asprenas, M. Raius Rufus ou C Vtius[54]), pour en faire celui du plomb qu'elle produisait elle-même ? Il est difficile de le dire. Les lingots de Comacchio témoigneraient donc plutôt d'une production alors désorganisée. Les mines du célèbre district sont en crise, la production de plomb a chuté, et l'exportation s'en ressent[55] : le temps n'est plus où des producteurs comme les Pontilieni pouvaient expédier vers Rome plus de 600 lingots en une seule fois[56]. J.M. Abascal Palazón[57] a bien souligné les changements provoqués dans la société de *Carthago Noua* par l'évolution des mines de plomb-argent. Les lingots de Comacchio en découvrent un nouvel aspect.

Agrippa et L. Cae(…) Bat(…) sont au centre de l'opération commerciale révélée par l'étude des timbres. Ce sont eux qui ont rassemblé les 102 lingots. Ceux qui leur ont fourni la plupart de ces derniers s'appellent Furii, Iunii, Planii Russini ou encore Vtii. Ils ont des noms bien connus à *Carthago Noua*. Naguère ceux qui s'appelaient ainsi étaient de grands producteurs de lingots, ils faisaient figurer orgueilleusement ces noms dans les estampilles moulées qui ornaient le dos des lingots. Ceux qui les portent à l'époque d'Agrippa ne sont plus que des marchands de plomb, qui apposent leurs timbres discrets sur tout ou partie des lingots qu'ils manient et qu'ils se sont procurés auprès d'artisans fondeurs qui travaillent, eux, vraisemblablement pour une clientèle locale, un plomb qui ne provient peut-être pas de la zone centrale de la Sierra[58], un plomb dont on ne sait dans quelles conditions il est produit et qui, de toute façon, n'est plus commercialisé sous la forme des beaux lingots évoqués ci-dessus. Ces artisans n'utilisent plus de matrices comme celles qui permettaient autrefois de produire les lingots standardisés de ce type, qui faisaient le renom de *Carthago Noua*[59]. Ils se contentent de modeler de leurs mains des lingotières dans le sable de fonderie en veillant à ce qu'elles soient d'un volume à peu près égal (séries I et II), ou bien ils utilisent comme matrices des billes de bois à peine équarries (série III). Au mieux, certains ont fabriqué des matrices plus régulières, de forme parallélépipédique (séries IV et V), différentes des modèles anciens par leur section trapézoïdale et non parabolique, et ils ont modelé dans le cartouche dorsal un caducée, un des symboles qui ornaient parfois l'estampille des lingots d'autrefois. Ainsi ont été faits ces lingots hors norme qui vont constituer le stock acquis par Agrippa et son associé, soit qu'ils les aient achetés selon des procédures diverses[60] à ces négociants en plomb de *Carthago Noua* qui ont nom Furii, Vtii, Iunii et Planii Russini, soit qu'ils se les soient procurés directement (série V). Précisément, on se rappellera que les signatures isotopiques du plomb de trois lingots de cette série étaient toutes proches de celle d'une mine de Mazarrón (fig. 6) : on pourrait donc formuler l'hypothèse que, ne trouvant pas assez de plomb auprès des marchands de *Carthago Noua*, L. Cae(sius ou -dius ?) Bat(ialus ?) soit allé s'en procurer, sans intermédiaire, dans les mines voisines de Mazarrón. L'hypothèse est simplement envisageable, mais les résultats des analyses isotopiques du plomb paraissent l'autoriser.

---

[51] Strabon semble en faire une loi générale, valable pour toutes les mines d'argent (et de plomb), qui, écrit-il, existent toujours mais ne sont plus du domaine public : elles sont devenues propriétés de particuliers. Pourtant, il semble bien qu'alors, les mines d'argent de Linares-La Carolina par exemple continuaient à faire partie du domaine public. Mais sans doute ce changement de statut a dû nécessiter du temps. En tout cas, il s'agirait alors d'une décision générale, à coloration politique.
[52] Orejas & Antolino, 1999, II 1A ; Orejas & Ramallo 2005, p. 97-99.
[53] Domergue 1990, p. 257 (n° 1026).
[54] Domergue 1990, p. 255-256 (n° 1025, 1037, 1054).
[55] Sur la fin de la grande production des mines de *Carthago Noua*, voir Domergue 1990, p. 230-234. La céramique recueillie dans les déblais miniers antiques, au cœur de la Sierra de Cartagena, confirme la nette baisse d'activité vers cette période : cf. Domergue 1987 b, p. 373-380, 383-385, 387-389.
[56] Épave de Mal di Ventre (Sardaigne) cf. Salvi 1992, p. 663.
[57] Abascal Palazón 2002, p. 36-37.

Nous avons attribué aux personnages que représentent les timbres 3 (GEME), 4 (MAC), 5 (C. MATI) et 6 (MAT) d'autres rôles. Le premier, Geme(llus ?), a son timbre présent sur un assez grand nombre de lingots des séries I à II ; par ailleurs, il ne saurait être assimilé à un fournisseur de base comme les Furii, Iunii, Vtii ou Plani Russini: son timbre est lié à celui d'Agrippa[61], alors que celui des Planii Russini fait partie des premiers imprimés (tableau 22, ligne 6), ceux précisément des fournisseurs de base. Gemellus a donc partie liée avec le duo Agrippa/L. Cae(…) Bat(…), ou à tout le moins travaille pour eux : en tout cas, il a le droit de marquer à son nom les lingots qu'il a achetés aux Furii et aux Nonii (il en est donc à ce moment le propriétaire) avant de les livrer aux deux associés.

---

[58] Voir ci-dessus, note 49.

[59] Domergue, Liou 1997, p. 11-20. Ces estampilles moulées, outre leur fonction publicitaire, répondaient peut-être à exigence à la fois commerciale et fiscale : celle d'un contrôle de la société publicaine chargée d'établir l'assiette du prélèvement qu'elle faisait sur les produits des mines du domaine public, et d'où elles tiraient à la fois le montant de la somme due à l'État romain pour la ferme des impôts et son propre bénéfice. Une fois ce système évanoui, ces estampilles devenaient inutiles.

[60] À propos des Planii Russini, une précision s'impose : le timbre PLR figure non seulement sur tous les lingots de la série IV, mais aussi sur quelques-uns des séries I à III, qui n'en portent aucun autre que le 7 (les timbres 1 et 2 mis à part, bien sûr). Nous avons admis plus haut que les marchands de lingots pouvaient ne pas imprimer leur timbre sur tous les exemplaires qui passaient entre leurs mains. Il ne serait donc pas illégitime de penser que les lingots des séries I à III marqués par le timbre 7 pouvaient faire partie, à l'origine, du stock fourni par les Vtii (les deux lingots porteurs du timbre VTI n'en ont effectivement pas d'autre que ce dernier), ce qui signifierait que les Planii Russini auraient pu racheter le stock des Vtii avant de le revendre à Agrippa et L. Cae(…) Bat(…). Dans ces conditions, les Planii auraient pu jouer les deux rôles : fournisseurs de base (série IV) et intermédiaires entre les Vtii et les deux associés. Mais on peut aussi considérer que les Planii Russini ont simplement ajouté à leur stock de la série IV quelques exemplaires supplémentaires appartenant aux séries I à III..

D'une façon générale, il faudrait disposer de plus de lingots (en particulier d'exemplaires porteurs des timbres les moins fréquents) pour pouvoir mieux analyser ces pratiques commerciales. D'ailleurs on peut penser que le stock finalement constitué par M. Agrippa et L. Cae(…) Bat(…) était plus important et comportait d'autres lingots, lesquels auront été transportés par d'autres bateaux.

[61] Voir *supra* le tableau de la figure 13 et le commentaire *k*.

Le rôle du dernier personnage, C. Matius (n° 5), représenté par le timbre 5 et sans doute aussi par les n° 4 et 6, est tout à fait différent. C'est vraisemblablement en effet le dernier propriétaire du stock de lingots, à qui Agrippa et L. Cae(…) Bat(…) les ont vendus. Ce qui peut pourtant surprendre, c'est d'abord le si petit nombre de lingots, 14 au total seulement, qui affichent les timbres de ce personnage, tandis qu'aucun des lingots des séries III, IV et V ne les porte, ni ceux des séries I et II[62] sur lesquels apparaissent les timbres des Furii, Planii Russini, Junii ou Vtii. Comme si C. Matius n'avait pas été le propriétaire de tous les lingots de ce stock. Par ailleurs, les cas de superposition des timbres (fig. 14) montrent que les derniers timbres à avoir été imprimés sont bien ceux de C. Matius. Comment concilier ces données contradictoires ? Peut-être en essayant de se représenter de quelle façon, concrètement, pouvaient s'être passées les choses. Imaginons donc le tas de lingots qu'Agrippa et L. Cae(…) Bat(…) viennent de vendre à C. Matius. Ce dernier le fait marquer avant de l'embarquer. Ses agents s'en chargent, mais, au lieu de déplacer les 102 lingots pour tous les marquer, ils vont se contenter de le faire sur ceux qui sont sur le dessus du tas. Il se trouve que ce sont des lingots qui n'ont pas d'autre timbre, sans doute parce que les autres, placés au-dessous d'eux, étaient restés plus ou moins groupés par lots, selon les divers fournisseurs.

La découverte de cette cargaison de lingots de plomb dans l'Adriatique nord (Fig. 15), n'a pas de quoi surprendre. Une liaison directe semble en effet avoir existé entre l'Adriatique et *Carthago Noua*, comme le laissent penser d'une part les épaves à amphores Lamboglia 2 présentes dans le voisinage de la Nouvelle-Carthage (Punta de Algas, Escombreras II) ou sur l'itinéraire qui y conduisait (San Jordi, Majorque), d'autre part la relative abondance de ces amphores dans la cité elle-même et dans les mines d'alentour[63]. Au retour, les bateaux pouvaient embarquer des cargaisons de lingots de plomb destinées à l'Adriatique-nord, comme le montrent la présence d'un lingot originaire des mines de *Carthago Noua*[64] au Magdalensberg (où par ailleurs les canalisations de distribution d'eau sont faites de ce plomb), et la découverte d'un second près de l'itinéraire qui y conduit, dans le lit de la rivière la Stella, un peu à l'ouest d'Aquilée[65].

L'itinéraire direct qui vient d'être évoqué supposerait que C. Matius soit entré en possession de la cargaison de lingots à *Carthago Noua* même. Mais cela cadre assez mal avec l'image qui nous paraissait vraisemblable d'un Agrippa cherchant à se procurer auprès des négociants locaux un plomb difficile à trouver : pourquoi, après l'avoir rassemblé, le revendre aussitôt? D'autant que les intérêts d'Agrippa sont à Rome et en Italie, pas en Espagne.

Or il y a un second scénario possible, c'est qu'Agrippa ait revendu à C. Matius le stock de lingots non pas en Espagne, avant qu'ils ne soient embarqués pour l'Italie, mais en Italie même, une fois la cargaison arrivée à bon port, par exemple à *Puteoli*, le port habituel de Rome à cette époque. C'est là que C. Matius serait entré en possession des lingots, et c'est de là qu'il les aurait envoyés vers l'Adriatique. Si Agrippa travaille pour son propre compte, il n'y a rien là que de normal : la revente est le fondement du commerce. En revanche, si cette cargaison a un caractère public, ou semi-public, dans la mesure où, en quelque sorte, Agrippa était alors chargé de l'urbanisme et de l'alimentation en eau de Rome, on pourrait s'étonner qu'elle ait été découverte si loin de l'*Vrbs*. Mais, en réalité, la revente de plomb par un service public paraît avoir été tout à fait possible. C'est par exemple ce qui semble être arrivé au lingot estampillé P. AEMILI. GALLICI trouvé dans une maison de Pompéi et qui était frappé du timbre NER AVG (Nero Augustus)[66], et peut-être à l'un des deux lingots mis au jour dans un atelier de plombier d'Herculanum, qui porte le timbre AVC ou AVG[67]. Tel a pu être aussi le sort de la cargaison de plomb de Comacchio.

---

[62] À l'exception d'un seul, le lingot 25, qui porte à la fois sur le dos le timbre des Iunii et celui de C. Matius. Mais ce lingot pose un problème : dans le catalogue de Berti 1990, p. 176, seul le timbre C. MATI est mentionné, mais sur le dessin de la p. 179, le timbre 9a apparaît clairement. Pour notre part, en 1982, nous n'avions rien vu sur le dos de ce lingot, mais ce dernier a pu être nettoyé entre 1982 et 1990, ce qui aurait fait apparaître les deux timbres.
[63] Molina 1997 ; Pérez 1998, p. 256 ; Pérez & Pascual 2004.
[64] Estampillé C. Iunius. Paetus (déjà mentionné supra, p. 000).
[65] Vitri et al. 1994. Ce lingot porte l'estampille moulée de C. Vtius.

Considérons maintenant l'épave et le lieu de sa découverte. Il s'agit d'un bateau de petit tonnage (21 m de longueur, 5,62 de large), au fond presque plat, apte à la fois au cabotage et à un trafic fluvial[68], mais un peu moins à la navigation hauturière. D'autre part, comme le suggère la position de l'épave dans la reconstitution du cadre paléo-géomorphologique du naufrage[69], le bateau devait longer la côte. Dans quel sens : nord-sud ou sud-nord ? L'une et l'autre direction sont possibles. On a eu jusqu'ici tendance à privilégier la deuxième[70], le bateau venant de Ravenne, où il aurait chargé des marchandises d'origines diverses, parmi lesquelles les lingots de plomb, pour les redistribuer dans la vallée du Pô, vers l'un des bras duquel il se dirigeait. Mais le sens nord-sud est tout à fait possible. Dans ce cas, le bateau viendrait d'Aquilée[71], le grand port de l'Adriatique-nord, en relation avec l'Istrie, l'Italie, l'*Hinterland* alpin (régions du Norique et de la Pannonie, et au-delà), l'Orient, l'Égypte et l'ensemble du monde méditerranéen[72]. Les diverses marchandises dont était constituée sa cargaison[73] - produits locaux ou régionaux : les amphores (de vin ? d'huile ?) Dressel 6A, les céramiques fines des types « Aco et « Surius », peut-être même les *naiskoi* en plomb : mais aussi produits importés : amphores de vins de Chios, de Cos et de Cnide, enfin bien sûr le stock de lingots - correspondent bien à ce qu'un port comme Aquilée pouvait recevoir et redistribuer. Dans cette hypothèse, soit le bateau aurait eu la même destination que dans la solution précédente, soit il aurait fait naufrage en se dirigeant vers Ravenne, qui aurait été sa vraie destination finale[74].

Aquilée était d'ailleurs le point d'aboutissement normal des lignes de navigation hauturière se dirigeant vers le nord de l'Adriatique, en longeant plutôt le rivage illyrien que la côte italienne[75]. Dans ces conditions, le vaisseau qui amenait les lingots de *Carthago Noua*, soit directement, soit via *Puteoli*, serait passé par le détroit de Messine, aurait contourné la botte et aurait navigué dans l'Adriatique jusqu'à Aquilée. Là les lingots auraient été débarqués, puis chargés sur une embarcation plus petite, celle qui a fait naufrage à Comacchio.

Avec un tel itinéraire, les lingots de Comacchio illustrent un des modèles habituels du commerce maritime antique[76], fondé d'une part sur un commerce à grande distance entre « places spécialisées » - de *Carthago Noua* à Aquilée (soit directement, soit via *Puteoli*) - d'autre part sur un commerce de redistribution, dirigé, à partir de l'emporion d'arrivée, Aquilée, soit vers la plaine Padane, soit sur Ravenne.

L'hétérogénéité et le caractère artisanal des lingots de Comacchio, l'absence d'estampille nominale moulée, l'intervention de personnages portant les noms de familles minières connues à *Carthago Noua* mais agissant en tant que négociants en plomb et non comme producteurs, autant de circonstances qui nous ont paru montrer que se procurer du plomb de commerce dans cette cité à la fin du I$^{er}$ siècle avant notre ère était une tâche moins facile que quelque cinquante ou soixante ans auparavant. Aussi n'est-il pas surprenant qu'Agrippa se soit aussi tourné vers celles des mines d'Espagne qui sont alors en pleine activité et produisent pour l'exportation dans les mêmes conditions que naguère à *Carthago Noua*, à savoir celles de la Sierra Morena. C'est ce que montre un lingot qui provient d'une épave trouvée dans les eaux de l'île de Minorque ; il a été produit par une *Soc(ietas) Pl(umbaria)*, nommée dans l'estampille moulée, et il porte le timbre AGRIP; le matériel d'accompagnement date le naufrage entre 50 avant et 50 après J.-C. : la fourchette est large, mais la présence du timbre AGRIP et le rapprochement avec les lingots de Comacchio devraient permettre de la réduire aux dernières décennies du Ier siècle avant J.-C.. D'après l'analyse isotopique du plomb, ce lingot doit provenir des mines du district de Linares-La Carolina (Jaén)[77]. Il s'agit d'un exemplaire de type classique, de belle forme et muni d'une estampille moulée. Le timbre AGRIP, qui, par sa facture, rappelle celui des lingots de Comacchio, est le seul présent. Ce lingot était une marchandise d'exportation courante : le représentant d'Agrippa l'aura acheté[78] sans intermédiaire à un producteur de l'est de la Sierra Morena (la *Societas Plumbaria*).

---

[66] Domergue 1990, p. 257, n° 4001, et p. 273.
[67] Monteix 2004, p. 367-368.
[68] Bonino 1990, p. 39.
[69] Bondesan *et al.* 1990, p. 20-21.
[70] Domergue 1987, p. 136 ; Zerbini 2002, p. 826.
[71] Arnaud 2005, p. 112.
[72] Sotinel 2002
[73] Berti 1990 b.
[74] Dans ce cas le plomb n'aurait-il pas pu être destiné à la flotte installée par Auguste à Ravenne (Suétone, *Diuus Augustus*, 49)? Le stock de lingots aurait alors conservé son caractère public ou semi-public, C. Matius pourrait avoir fait partie du *staff* d'Agrippa et le stock de lingots aurait pu partir de *Carthago Noua* directement pour Aquilée, sans que l'on eût à imaginer un transbordement à *Puteoli* : autant d'hypothèses qui s'enchaînent mais qui sont difficilement vérifiables !
[75] Arnaud 2005, p. 193-205.
[76] Arnaud 2005, p. 107-126.

<div style="text-align:center">+<br>+ +</div>

Conclusion

Dans cette étude, l'archéométrie a joué un grand rôle, en raison du recours aux analyses isotopiques du plomb. À l'heure actuelle, cette méthode est sans doute la plus sûre pour déterminer l'origine d'un métal, à condition que ce métal renferme du plomb à des teneurs suffisantes pour permettre son utilisation. En plus du plomb bien sûr, c'est le cas du cuivre, de l'argent, de l'or et, dans une certaine mesure, de l'étain. C'est plus problématique pour le fer. En tout cas, c'est un genre d'analyse qui s'applique avec de bonnes chances de succès à des marchandises comme des lingots, produits semi-finis destinés à être commercialisés. À ce stade en effet, il y a peu de chances qu'on ait affaire à des mélanges. Les conditions sont encore meilleures lorsqu'il s'agit de lingots de plomb, comme ici.

L'analyse de ceux de Comacchio constitue en effet un excellent exemple de ce genre d'études. Comme on le sait, les mesures isotopiques du plomb et l'utilisation d'une banque de données correspondantes permettent avant tout, lorsqu'on s'intéresse à l'origine d'un produit, d'écarter un certain nombre d'hypothèses, parce que les signatures isotopiques s'y opposent, pour ne laisser ouvertes que quelques possibilités. Ici, finalement, il n'en est resté que deux, l'une et l'autre dans le sud-est de l'Espagne : la Sierra Almagrera et le district de Cartagena-Mazarrón. Mais, pour en arriver à ce résultat, les disciplines historico-archéologiques avaient dû intervenir pour écarter, elles aussi, des districts miniers dont la signature isotopique était compatible avec celle des lingots, mais qui n'avaient pas été exploités à l'époque considérée : c'est le cas de la Toscane, et surtout du Cabo de Gata. Mais comment choisir entre la Sierra Almagrera et Cartagena-Mazarrón? L'épigraphie a permis de le faire, en montrant que les noms qu'on pouvait lire sur les timbres portés par les lingots étaient ceux de familles bien connues à *Carthago Noua* à l'époque antérieure comme productrices de plomb. La Sierra Almagrera se trouvait donc écartée, et les mines de Carthagène apparaissaient comme étant celles d'où provenaient les lingots. Mais ici encore, les isotopes du plomb s'invitaient une nouvelle fois au débat, en suggérant que les mines de Mazarrón, toutes proches, avaient pu, elles aussi, fournir une petite partie (la série V) des lingots de Comacchio.

---

[77] Analyses P.R. Trincherini, JRS, Environment Institute, CCR, Ispra, Varese. Voir aussi Rodà 2003.
[78] Cet exemplaire faisait partie d'un lot plus important, mais l'épave a été pillée à plusieurs reprises, ce qui a causé la perte des autres exemplaires (Rodà 2003, p. 184).

On voit donc que, s'agissant de la provenance des métaux antiques, les données isotopiques du plomb ne permettent pas, à elles seules, de donner une réponse définitive : les méthodes de laboratoire et les disciplines archéologico-historiques s'appuient mutuellement pour permettre une approche mieux orientée d'une réalité toujours difficile à cerner. En tout cas elles ne peuvent s'ignorer ni se mépriser mutuellement. Elles sont destinées à collaborer.

Dans le cas présent, la détermination de la région d'origine des lingots a permis de replacer ces derniers dans leur contexte historique, celui d'une deuxième phase de production des mines de *Carthago Noua*, après une première phase plus brillante, qui, à partir du milieu du II$^e$ siècle avant J.-C. et jusqu'au milieu du siècle suivant, avait vu les produits de ces mines, bien identifiables en particulier par les noms de leurs fabricants, dominer le marché méditerranéen occidental. Il semble que, dans l'avant-dernière décennie du I$^{er}$ siècle avant J.-C., la situation ait été moins favorable : alors, comme l'indique Strabon, les mines sont devenues propriétés privées, ce qui semble correspondre à une baisse de la production, ou, tout au moins, à un changement dans le mode d'exploitation, qui, peut-être, rendent le commerce du plomb difficile. C'est à ce moment-là qu'Agrippa, *patronus* de *Carthago Noua*, a cherché à s'en procurer dans ces mines ; après s'être associé avec un certain L. Cae(sius ou –dius) Bat(ilius ?), sans doute une personnalité locale, et avec l'aide d'un nommé Gemellus, à la fois intermédiaire dans ce commerce et agent des deux associés, il a pu réunir une assez grande quantité de lingots que lui ont fournis plusieurs marchands descendants des grandes « familles du plomb », les Furii, Iunii, Planii Russini, Vtii ; mais ce ne fut peut-être pas sans difficulté, comme le suggérerait la présence d'une série de lingots venus peut-être de Mazarrón (série V) pour compléter le stock.

Les responsabilités d'Agrippa dans l'urbanisme augustéen de Rome expliquent que le gendre d'Auguste ait pu se faire ponctuellement négociant en plomb et qu'il s'en soit procuré tant auprès des mines de *Carthago Noua* que de celles du district de Linares-la Carolina, comme le montre un lingot frappé du timbre d'Agrippa, trouvé dans les eaux de Minorque.

Il s'agissait là sans doute de matières premières qui étaient propriétés publiques, mais que, dans certaines circonstances, les magistrats pouvaient aliéner à des particuliers. Agrippa et son associé ont pu le faire pour les 102 lingots de *Carthago Noua*, sans doute à *Puteoli*, en faveur d'un certain C. Matius Mac(edo ?), qui paraît bien avoir été leur dernier propriétaire. De là, les lingots ont été dirigés vers l'Adriatique-nord, sans doute vers le grand port d'Aquilée, où ils ont été débarqués. Avec d'autres marchandises de provenances diverses (Méditerranée orientale, Italie du nord, région Adriatique), ils ont été chargés sur une nouvelle embarcation, à destination de la plaine de Padane ou de Ravenne. Mais arrivé au voisinage du delta du Pô, à une date difficile à fixer entre 19 et 12 avant J.-C., le bateau a fait naufrage et s'est échoué près de la côte, en un point où, quelque vingt siècles plus tard, il sera découvert, conservé en pleine terre grâce aux alluvions du fleuve.

*Légendes des figures*

.
Figure 1 – Trois des cent-deux lingots de Comacchio (de droite à gauche : séries II, IV et I).
Figure 2 – Résultats des analyses isotopiques du plomb des lingots de Comacchio (analyses P.R. Trincherini, JRS, Environment Institute, CCR, Ispra, Varese). Le lingot Berti n° 69 a été analysé à deux reprises.
Figure 3 – Diagramme de l'âge : $^{206}Pb/^{204}Pb$ versus $^{207}Pb/^{204}Pb$ (A. Nesta et P. Quarati ).
Figure 4 - Diagramme de l'âge : $^{206}Pb/^{204}Pb$ versus $^{207}Pb/^{204}Pb$ (détail) (A. Nesta et P. Quarati ).
Figure 5 – Diagramme de la précision : $^{207}Pb/^{206}Pb$ versus $^{208}/^{206}Pb$ (A. Nesta et P. Quarati ).
Figure 6 – Diagramme de la précision : $^{207}Pb/^{206}Pb$ versus $^{208}/^{206}Pb$ (détail) (A. Nesta et P. Quarati).
Figure 7 - Diagramme de la précision : $^{207}Pb/^{206}Pb$ versus $^{208}/^{206}Pb$. Les districts miniers de plomb-argent du sud-est de l'Espagne et les lingots de Comacchio (A. Nesta et P. Quarati ).
Figure 8 - Diagramme de la précision : $^{207}Pb/^{206}Pb$ versus $^{208}/^{206}Pb$. Détail du diagramme précédent (A. Nesta et P. Quarati ).
Figure 9 - Diagramme de la précision : $^{207}Pb/^{206}Pb$ versus $^{208}/^{206}Pb$. Les lingots de Comacchio et les principaux districts d'Hispanie producteurs de plomb dans l'Antiquité romaine (version « points ») (A. Nesta et P. Quarati ).
Figure 10 - Diagramme de la précision : $^{207}Pb/^{206}Pb$ versus $^{208}/^{206}Pb$. Les lingots de Comacchio et les principaux districts d'Hispanie producteurs de plomb dans l'Antiquité romaine (version « nuages») (A. Nesta et P. Quarati ).
Figure 11 – Un lingot de la série V (n° 98) et son estampille moulée représentant un caducée
Figure 12- Les timbres des lingots de Comacchio.
Figure 13 – La distribution des timbres. Ce tableau a été constitué à partir de deux séries de documents : d'abord le précieux catalogue publié par F. Berti (Berti 1990b) dans *Fortuna Maris* (Berti 1990a), où les lingots sont présentés un par un, chacun avec ses caractères techniques particuliers (mesures, poids), l'indication des timbres qu'il porte et de l'endroit où ils sont imprimés (base, dos, côté) : chacune de ces notices est en général accompagnée d'un dessin où sont figurés la base, le dos et les côtés du lingot et où sont reproduits les timbres à leur emplacement exact. Ce catalogue présente cependant un léger inconvénient : la même formule « sigla entro cartiglio » désigne indistinctement les timbres 7, 8 et 10, et les dessins, le plus souvent à petite échelle, ne permettent pas toujours de préciser. Aussi avons-nous eu également recours aux fiches rédigées par C.D. en mars 1982.
Nous conservons la numérotation de F. Berti (2$^e$ colonne ; entre parenthèses, les numéros de Domergue 1987) ainsi que la division en cinq séries (1ère colonne), qui est celle de son catalogue.
Dans la 2$^e$ colonne, les numéros soulignés sont ceux des lingots dont le plomb a été analysé.
L'emplacement des timbres est indiqué par les lettres B (base), D (dos), C (grand côté), c (petit côté)
Dans la dernière colonne, sont signalées les superpositions de timbres (cf. tableau de la figure 14).
Figure 14 – Les cas de superposition des timbres.
Figure 15 –Itinéraires probables :
1 ) des lingots de Comacchio, de *Carthago Noua* à Comacchio, via *Puteoli* et *Aquileia* (trait continu), avec une variante (en tireté : itinéraire direct) ;
2 - du lingot de Minorque, du district de Linares-la Carolina (Hispanie méridionale) à Minorque, via le *Baetis* (Guadalquivir) et les colonnes d'Hercule (pointillé).

Les étoiles indiquent les lieux de naufrage.

# Bibliographie

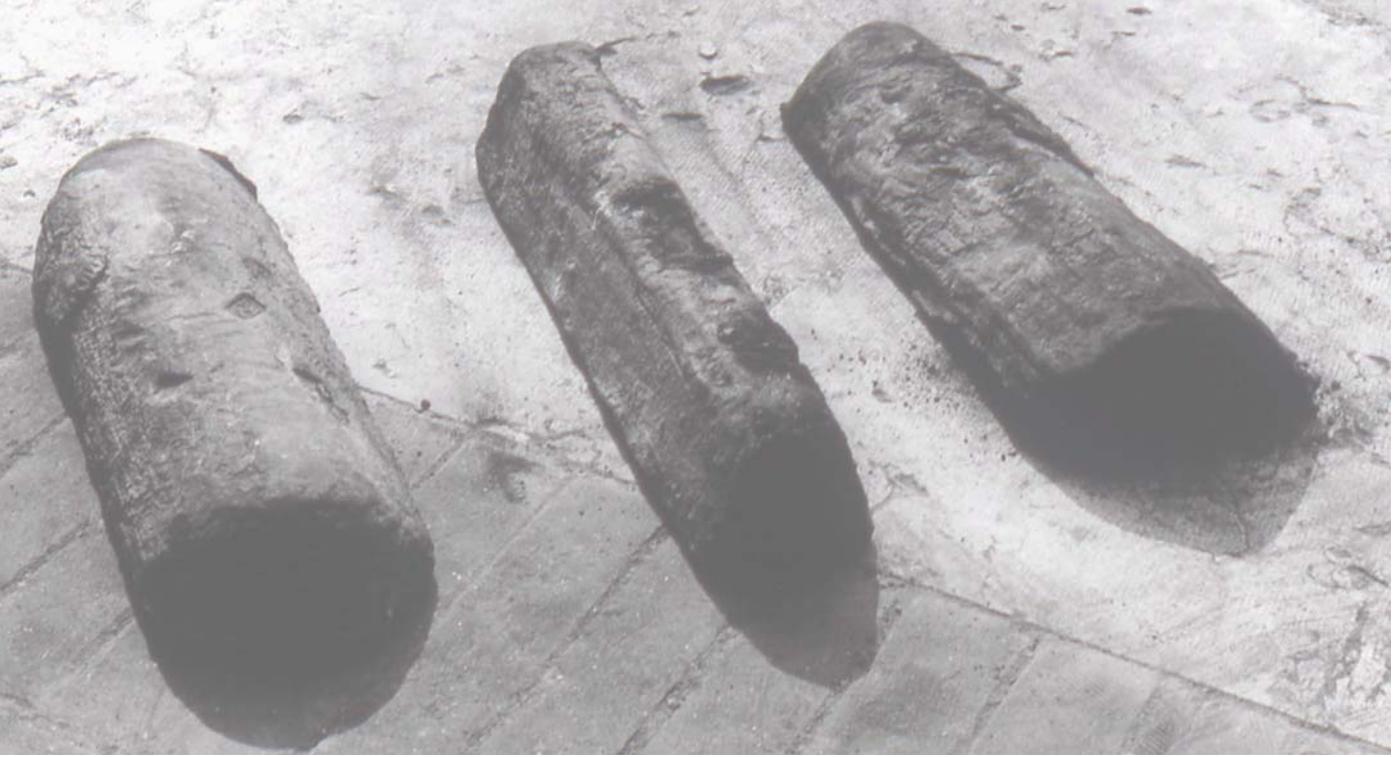

| Inv. Musée | Série | Catal. Berti (1990) | (6/4) | (7/4) | (8/4) | (6/7) | (7/6) | (8/6) |
|---|---|---|---|---|---|---|---|---|
| 54.864 | I | 3 | 18,676 | 15,65389 | 38,731 | 1,193058 | 0,838182 | 2,073838 |
| 54.859 | " | 4 | 18,692 | 15,668 | 38,801 | 1,193005 | 0,83822 | 2,075808 |
| 57.097 | " | 5 | 18,692 | 15,665 | 38,802 | 1,193233 | 0,838059 | 2,075861 |
| 57.093 | " | 8 | 18,677 | 15,659 | 38,727 | 1,192733 | 0,838411 | 2,073513 |
| 54.851 | " | 13 | 18,691 | 15,664 | 38,743 | 1,193246 | 0,83805 | 2,072816 |
| 54.854 | " | 15 | 18,685 | 15,668 | 38,786 | 1,192558 | 0,838534 | 2,075783 |
| 54.869 | II | 26 | 18,705 | 15,679 | 38,8292 | 1,192997 | 0,838225 | 2,075873 |
| 57.125 | " | 29 | 18,675 | 15,65 | 38,719 | 1,193291 | 0,838019 | 2,073307 |
| 57.126 | " | 41 | 18,66974 | 15,65146 | 38,72009 | 1,192843 | 0,838333 | 2,07395 |
| 57.065 | III | 43 | 18,681 | 15,659 | 38,739 | 1,192988 | 0,838231 | 2,073711 |
| 54.847 | IV | 68 | 18,6743 | 15,6497 | 38,725 | 1,193269 | 0,838034 | 2,073706 |
| 57.090 | " | 69 | 18,691 | 15,662 | 38,746 | 1,193398 | 0,837943 | 2,072976 |
| 57 090a | " | 69 | 18,67564 | 15,65067 | 38,7033 | 1,193281 | 0,838026 | 2,072395 |
| 54.839 | " | 71 | 18,669 | 15,6426 | 38,7059 | 1,193472 | 0,837892 | 2,073271 |
| 57.132 | " | 86 | 18,683 | 15,6572 | 38,742 | 1,193253 | 0,838045 | 2,07365 |
| 57.088 | " | 88 | 18,675 | 15,653 | 38,73 | 1,193062 | 0,838179 | 2,073896 |
| 57.122 | " | 89 | 18,678 | 15,653 | 38,739 | 1,193254 | 0,838045 | 2,074044 |
| 54.840 | " | 96 | 18,6739 | 15,65677 | 38,74 | 1,192704 | 0,838431 | 2,074553 |
| 54.835 | V | 99 | 18,787 | 15,644 | 38,822 | 1,200908 | 0,832703 | 2,066429 |
| 54.834 | " | 100 | 18,779 | 15,635 | 38,805 | 1,201087 | 0,832579 | 2,066404 |
| 57.106 | " | 102 | 18,79866 | 15,66121 | 38,82728 | 1,200333 | 0,833102 | 2,065428 |

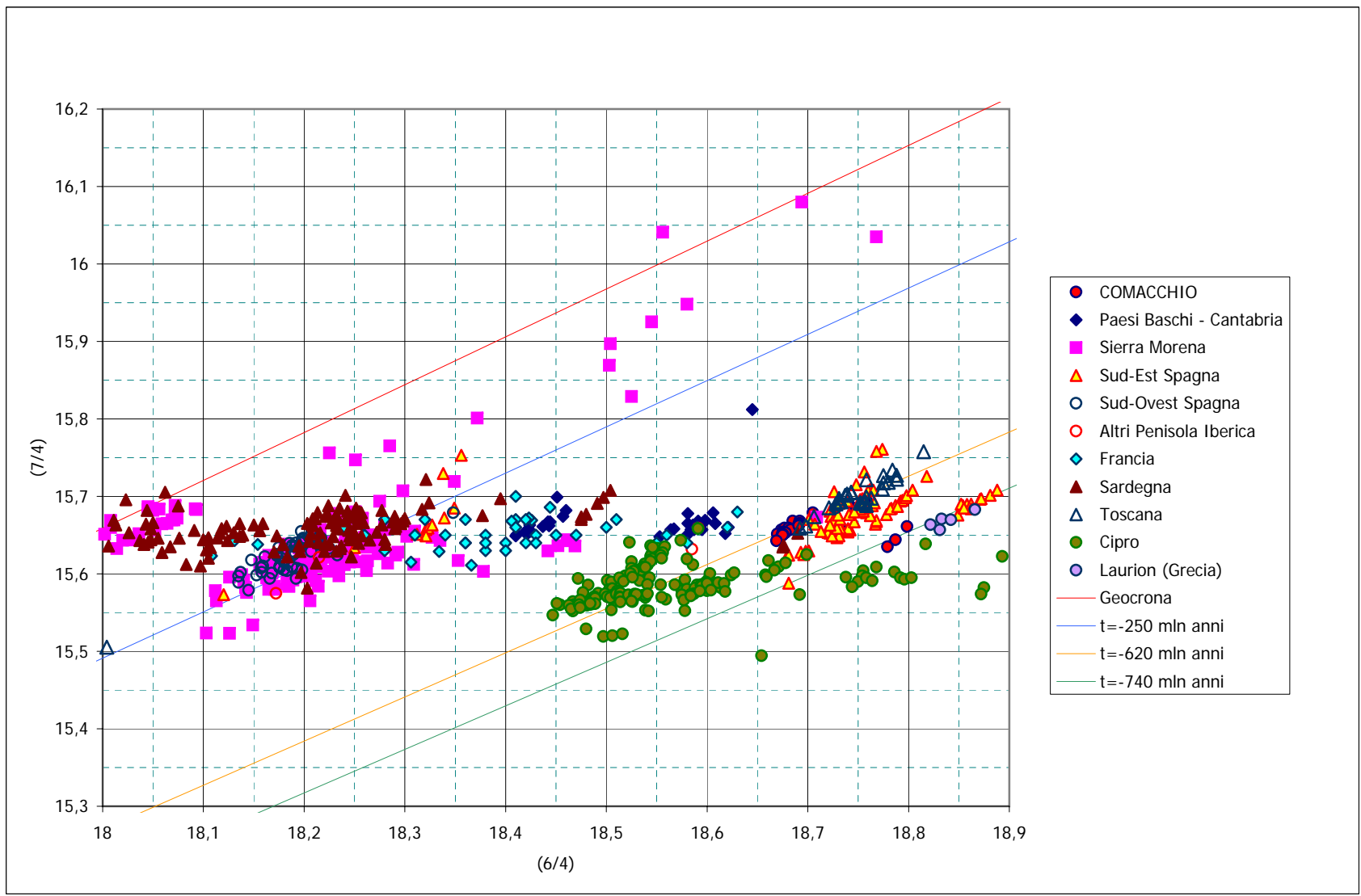

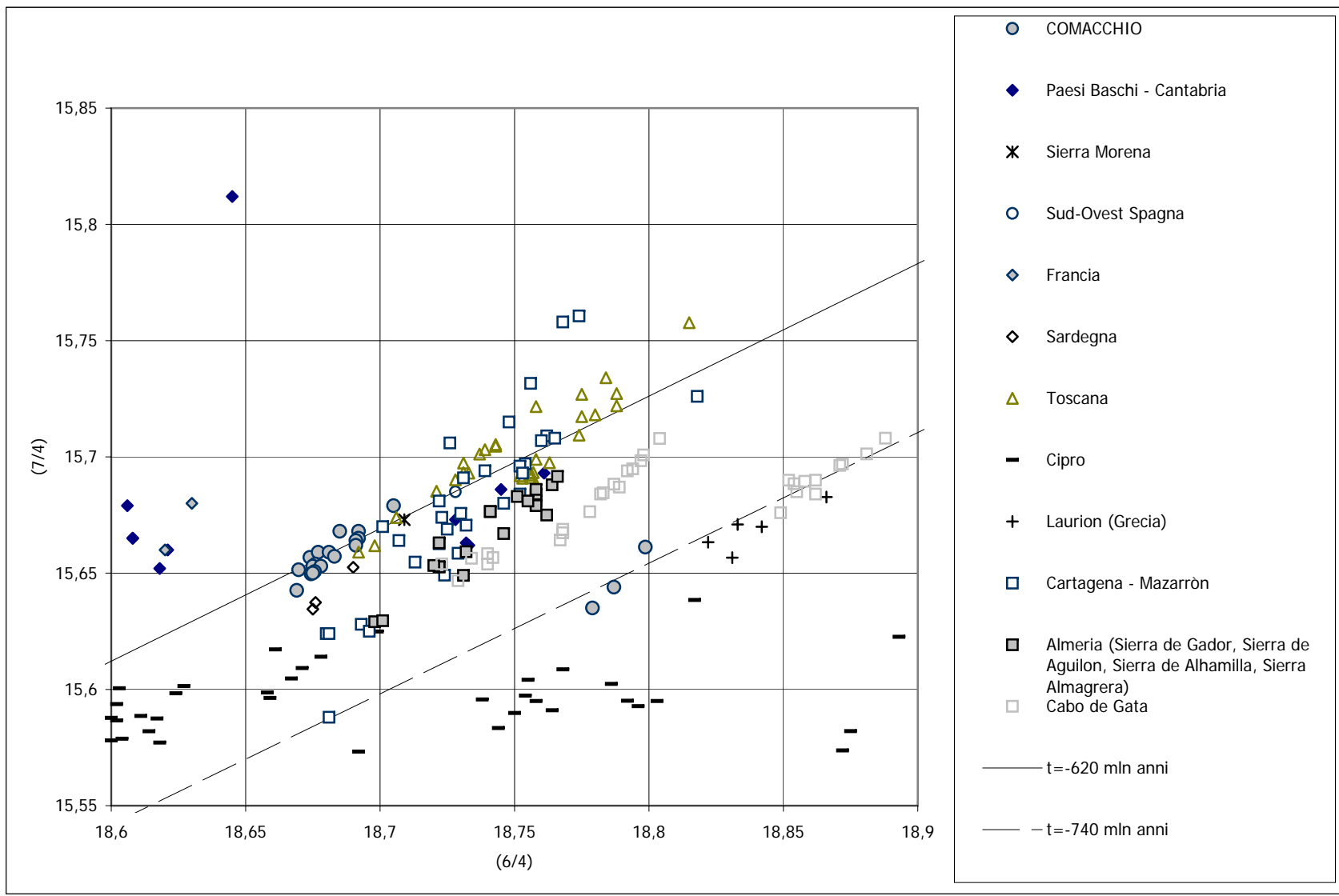

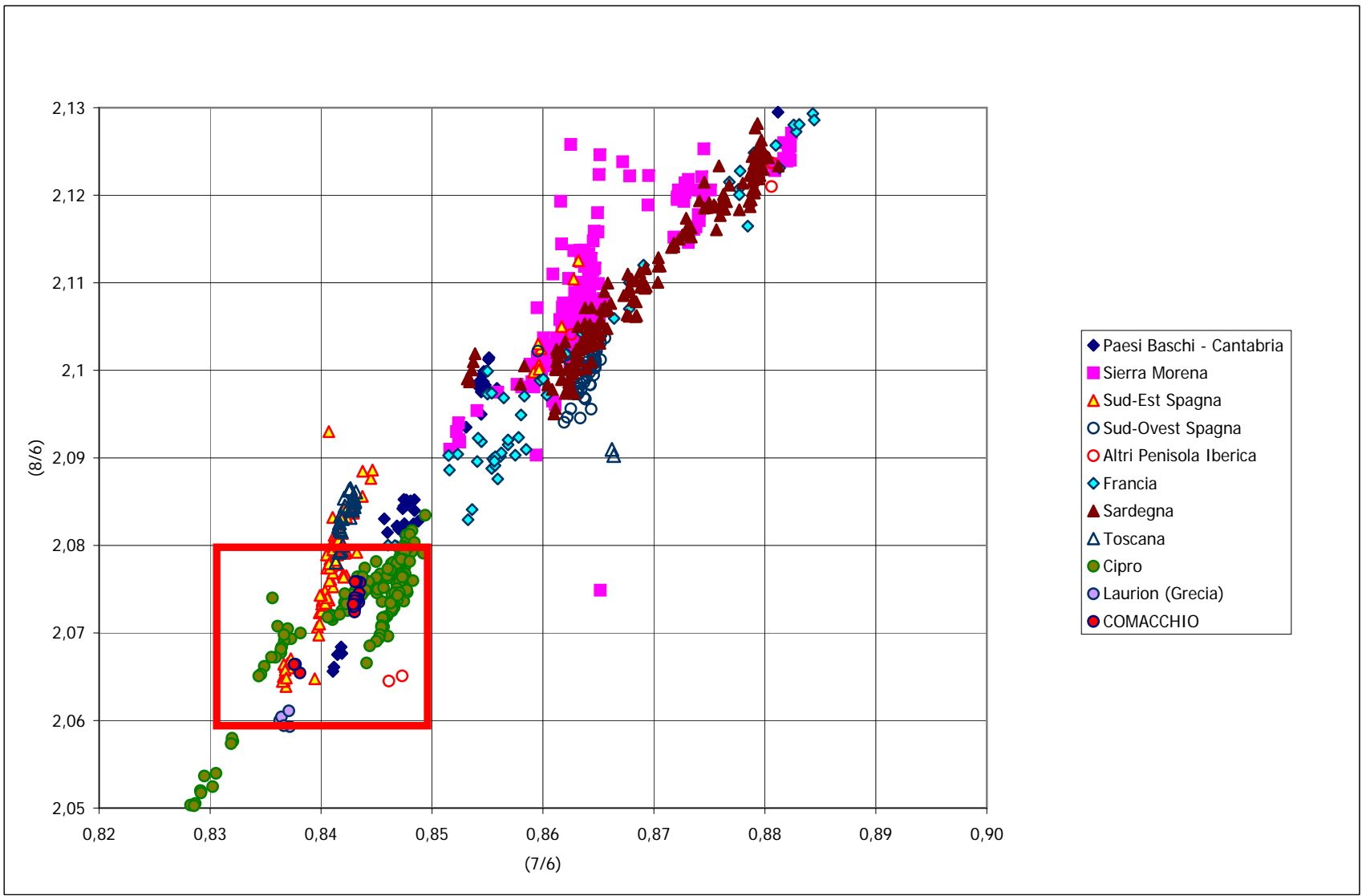

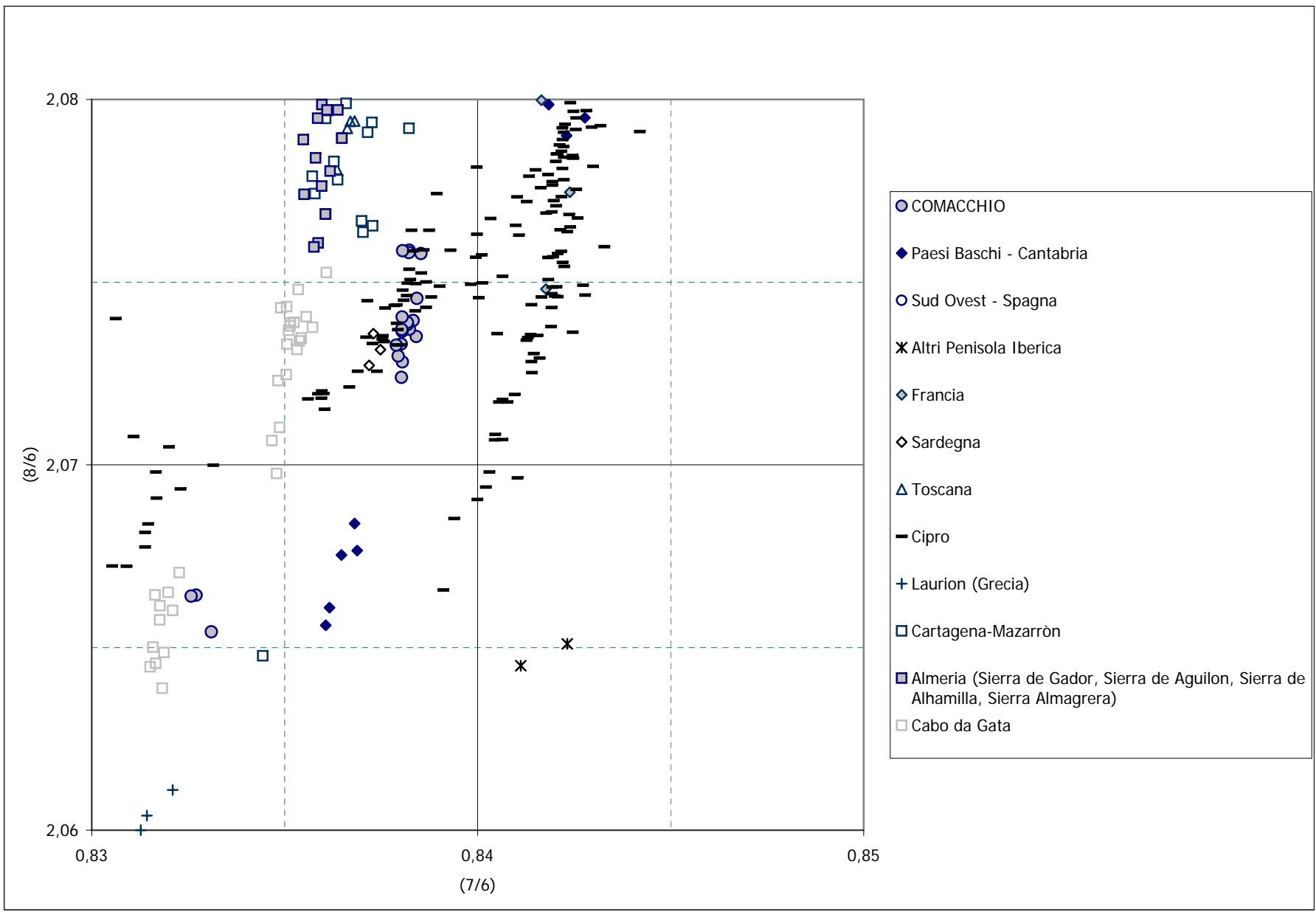

(8/6)

2,08

2,07

2,06

0,83　　　　　　　　　　　0,84　　　　　　　　　　　0,85
(7/6)

Mazarron

- ○ COMACCHIO
- □ Cartagena-Mazarròn
- ■ Almeria (Sierra de Gador, Sierra de Aguilon, Sierra de Alhamilla, Sierra Almagrera)
- △ Cabo da Gata

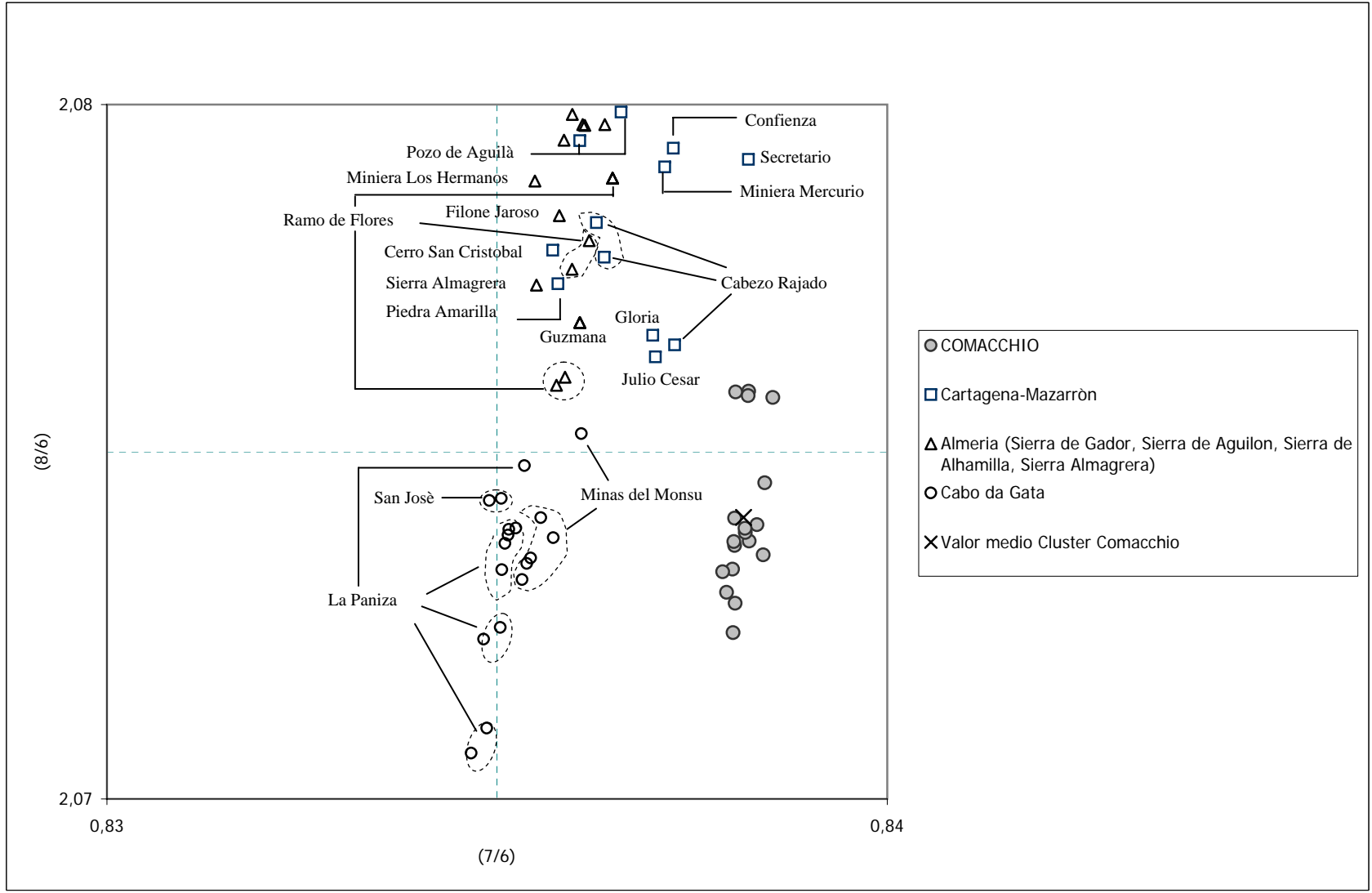

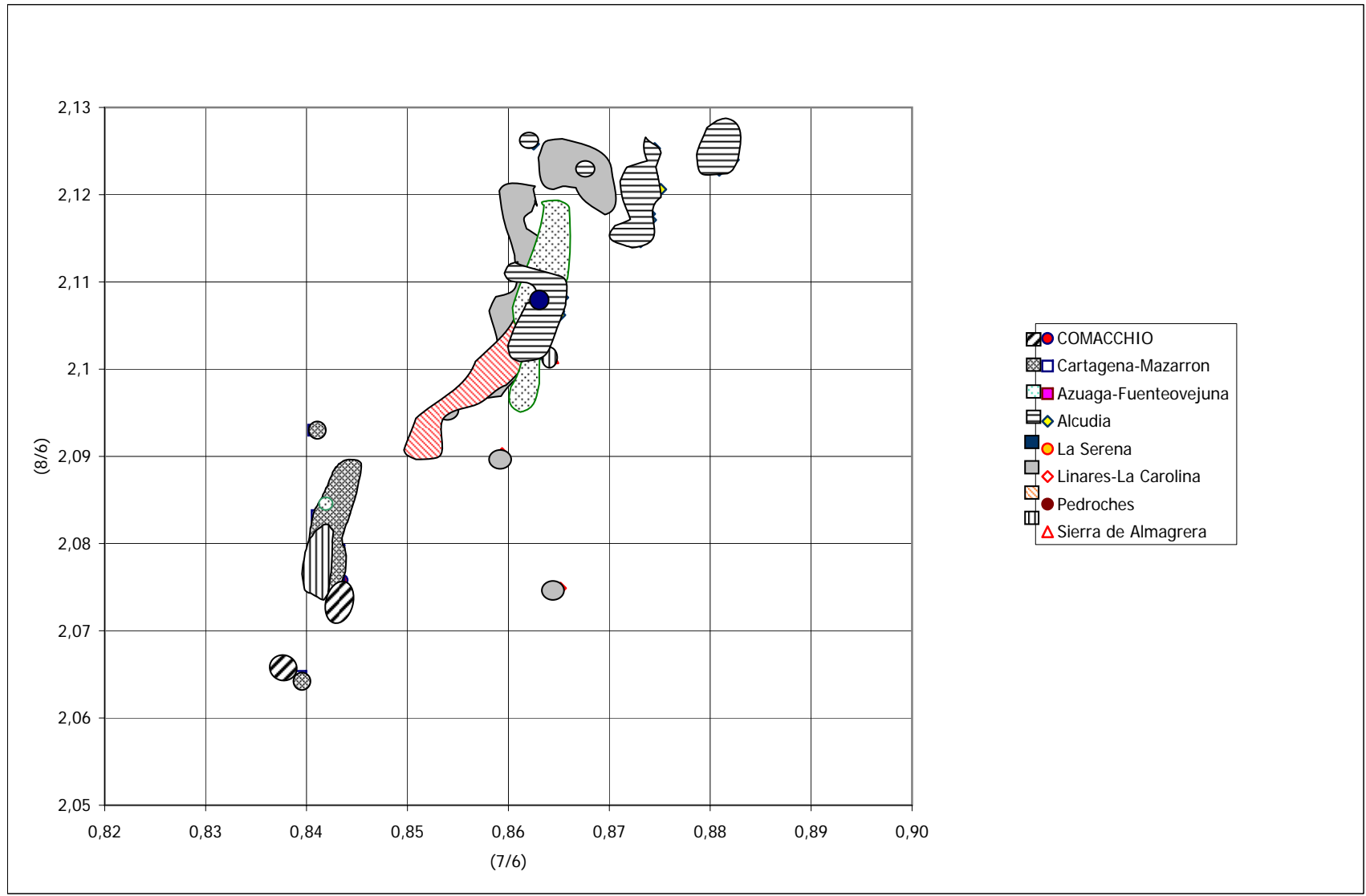

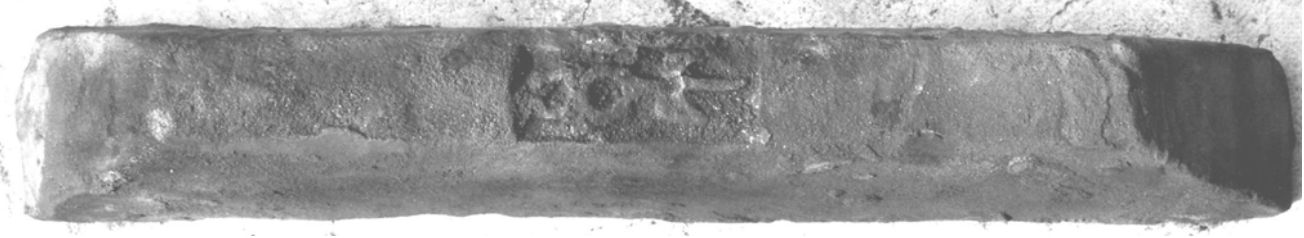

| | |
|---|---|
| ACRIP | L·CAE·BA |
| 1 | 2 |
| GEME | MAC |
| 3 | 4 |
| C·MATI | MAT |
| 5 | 6 |
| IRA | (monogram) |
| 8 | 7 |
| ⋈ | X |
| 9 | 9a |
| | (mark) |
| | 10 |

0 ———— 5 cm

| Série | Catalogue | Timbre 8 | Timbre 10 | Timbre 9 | Timbre 9 a | Timbre 7 | Timbre 3 | Timbre 1 | Timbre 2 | Timbre 5 | Timbre 4 | Timbre 6 | Graffito | Superposition |
|---|---|---|---|---|---|---|---|---|---|---|---|---|---|---|
| | | FVRI | VTI | IVNI | IVNI (?) | PI.R | GEME | AGRIP | L.CAE.BAT | C.MATI | MAC | MAT | sur la base | |
| I | 1 (1) | | | | | | | | | | | | | |
| - | 2 (29) | | | | | | B | B | B | | | | | 1>2 |
| - | 3 (62) | | | | | | | D | B | B | D | | | |
| - | 4 (57) | | | | | | | D | B | B | D | | | |
| - | 5 (14) | | | | | B | | B | B | | | | | |
| - | 6 (84) | | | | | | | D | B | | | | | |
| - | 7 (28) | | | | | | D | B | B | | | | | |
| - | 8 (24) | | | | | B | | B | B | | | | | |
| - | 9 (34) | | D | | | | | B | B | | | | | |
| - | 10 (36) | D | | | D | | D | D | B | | | | | |
| - | 11 (45) | | | | | | D | D | B | | D | | A | |
| - | 12 (75) | | | | | | | B | B | | | B | | |
| - | 13 | D | | | D | | D | B | B | | | | | |
| - | 14 (10) | | | | | | | D | B | | D | | | |
| - | 15 (52) | | | | | | | B | B | | | B | M | M >6>1 |
| - | 16 (59) | D | | | D | | D | B | B | | | | | |
| II | 17 (20) | | | | | | | D | B | B | D | | | |
| - | 18 (46) | D | | | D | | D | B | B | | | | | |
| - | 19 (43) | | D | | | | | B | B | | | | | |
| - | 20 (74) | | | | | | | | B | | D | | | |
| - | 21 (40) | | | | B | | | B | B | | | | | |
| - | 22 (5) | | | | | | D | B | | | | | | |
| - | 23 (33) | D | | | D | | D | B | B | | | | | |
| - | 24 (98) | | | | | | | B | | | | | | |
| - | 25 (96) | | | | D | | | B | B | D | | | | |
| - | 26 (53) | D | | | D | | D | B | B | | | | | |
| - | 27 (58) | | | | | B | | B | B | | | | | |
| - | 28 (18) | D | | | D | | | B | B | | | | | |
| - | 29 (71) | | | | | | | D | B | | | B | | |
| - | 30 (32) | | | | | | D | B | B | | | | | |
| - | 31 (4) | | | | | | | B | B | | | | | |
| - | 32 (90) | D | | D | | | D | B et D | B | | | | | |
| - | 33 (80) | D | | | D | | D | B | B | | | | | |
| - | 34 (42) | | | | | | B | B | B | | | | | 1>3 |
| - | 35 (92) | | | | | B | | B | B | | | | | |
| - | 36 (82) | | | | | | | B | B | | | | | |
| - | 37 (85) | | | | | | B | B | B | | | | | |
| - | 38 (69) | | | | | | B | D | B | | D | | | |
| - | 39 (68) | | | | | | | D | B | | D | | | |
| - | 40 (97) | | | | | | | B | B | | | | | |
| - | 41 (79) | | | | | | | D | B | | D | | | 4>1 |
| - | 42 (27) | | | | | | | B | B | | | | | |
| III | 43 | | | | | | | | B | | | | | |
| - | 44 (25) | | | | | | | B | B | | | | | |
| - | 45 (22) | | | | | | | D | B | | | | | |
| - | 46 (72) | | | | | | | B | B | | | | | 2>1 |
| - | 47 (73) | | | | | | | B | B | | | | | 2>1 |
| - | 48 (6) | | | | | | | B | B | | | | | 2>1 |
| - | 49 (91) | | | | | | | B | B | | B | | | |
| - | 50 (99) | | | | | | | B | B | | | | | 2>1 |
| - | 51 (19) | | | | | | B | B | B | | | | | |
| - | 52 (49) | | | | | | B | B | B | | | | | 3>1 |
| - | 53 (70) | | | | | | | | B | | D | | | |
| - | 54 (94) | | | | | | B | | B | | | | | |
| - | 55 (61) | | | | | | B | B | B | | | | NAB | |
| - | 56 (95) | | | | | B | | B | B | | | | | |
| IV | 57 (13) | | | | | B | | B | B | | | | | |
| - | 58 (21) | | | | | B | | B | B | | | | | |
| - | 59 (35) | | | | | B | | B | B | | | | | |
| - | 60 (3) | | | | | B | | | B | | | | | |
| - | 61 (38) | | | | | B | | | B | | | | | 2>7 |
| - | 62 (81) | | | | | B | | B | B | | | | | |
| - | 63 (86) | | | | | B | | C | B | | | | | |
| - | 64 (56) | | | | | B | | B | B | | | | | |
| - | 65 (11) | | | | | B | | c | | | | | | |
| - | 66 (83) | | | | | B | | B | | | | | | |
| - | 67 (55) | | | | | B | | C et B | B | | | | | |
| - | 68 (37) | | | | | B | | B | B | | | | | |
| - | 69 (77) | | | | | B | | B | B | | | | | |
| - | 70 (7) | | | | | B | | B | B | | | | | |
| - | 71 (16) | | | | | B | | C | B | | | | | |
| - | 72 (9) | | | | | B | | B | B | | | | | |
| - | 73 (67) | | | | | B | | B | B | | | | | |
| - | 74 (17) | | | | | B | | B | B | | | | | |
| - | 75 (87) | | | | | B | | B | B | | | | | |
| - | 76 (48) | | | | | B | | B | B | | | | | |
| - | 77 (60) | | | | | B | | B | B | | | | | |
| - | 78 (23) | | | | | B | | C | B | | | | | |
| - | 79 (65) | | | | | B | | B | B | | | | | |
| - | 80 (64) | | | | | B | | C | B | | | | | |
| - | 81 (63) | | | | | B | | B | B | | | | | |
| - | 82 (54) | | | | | B | | B | B | | | | | |
| - | 83 (66) | | | | | B | | B | B | | | | | |
| - | 84 (26) | | | | | B | | B | B | | | | | |
| - | 85 (31) | | | | | B | | B | B | | | | | |
| - | 86 | | | | | B | | C | B | | | | | |
| - | 87 (88) | | | | | B | | B | B | | | | | |
| - | 88 (2) | | | | | | | | | | | | | 2>1>7 |
| - | 89 (76) | | | | | B | | B | B | | | | | |
| - | 90 (93) | | | | | B | | | B | | | | | |
| - | 91 (89) | | | | | B | | C | B | | | | | |
| - | 92 (47) | | | | | B | | B | B | | | | | |
| - | 93 (44) | | | | | B | | C | B | | | | | |
| - | 94 (51) | | | | | B | | B | B | | | | | |
| - | 95 (41) | | | | | B | | C | B | | | | | |
| - | 96 (12) | | | | | B | | B | B | | | | | |
| - | 97 (8) | | | | | B | | B | B | | | | | |
| V | 98 (39) | | | | | | | | B | | | | | |
| - | 99 (30) | | | | | | | C | B | | | | | |
| - | 100 (50) | | | | | | | C | B | | | | | |
| - | 101 15 | | | | | | | C | B | | | | | |
| - | 102 (78) | | | | | | | C | B | | | | | |

| Cas n° | N° du lingot | Le timbre… | …est imprimé sur… | …lui-même imprimé sur |
|--------|--------------|------------|-------------------|----------------------|
| 1 | 2 | AGRIP | L.CAE.BAT | |
| 2 | 46, 47, 48, 50, 52 | L.CAE.BAT | AGRIP | |
| 3 | 34 | AGRIP | GEME | |
| 4 | 52 | GEME | AGRIP | |
| 5 | 61 | L.CAE.BAT | PLR | |
| 6 | 89 | L.CAE.BAT | AGRIP | PLR |
| 7 | 41 | MAC | AGRIP | |
| 8 | 15 | MAT | AGRIP | |

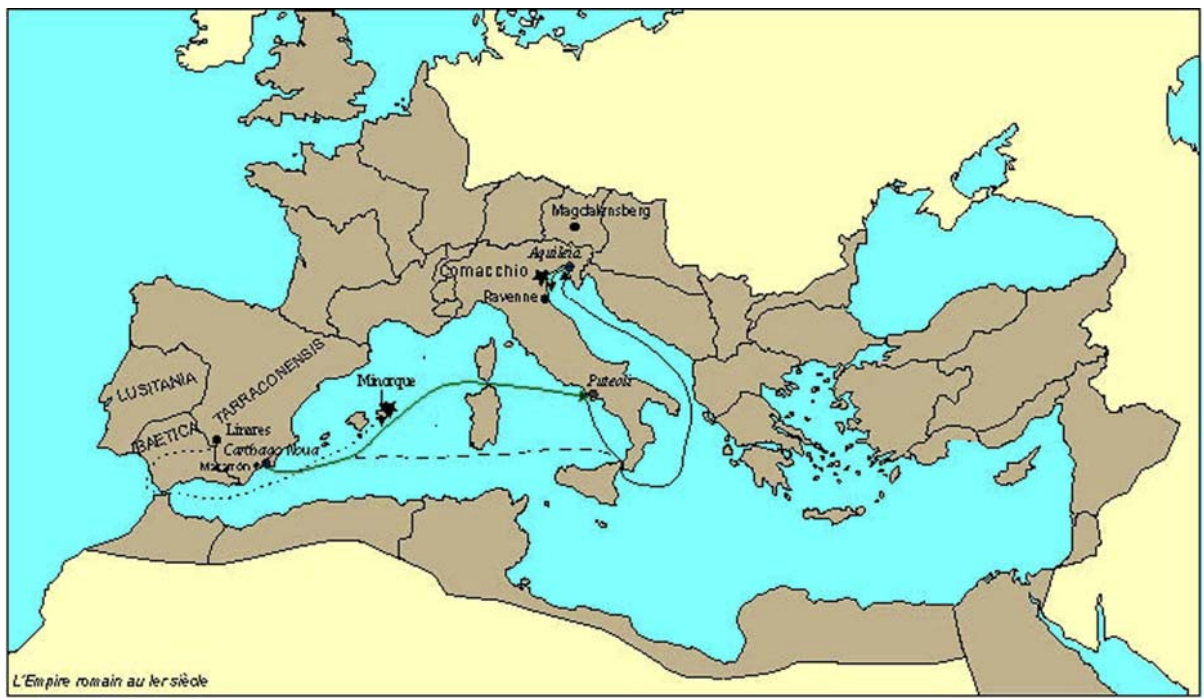